\documentclass{aastex}

\newcommand\hb{H${\beta}$~}
\newcommand\ha{H${\alpha}$~}
\newcommand\nii{[N{\sc ii}]~}
\newcommand\sii{[S{\sc ii}]~}
\newcommand\siii{[S{\sc iii}]~}
\newcommand\oii{[O{\sc ii}]~}
\newcommand\oi{[O{\sc i}]~}
\newcommand\hei{He{\sc i}~}
\newcommand\oiii{[O{\sc iii}]~}

\newcommand\sm{M$_{\odot}$}
\slugcomment{}
\shortauthors{Stanghellini et al.}
\shorttitle{Slitless spectroscopy} 

\begin{document}

\title{Slitless Spectroscopy of LMC Planetary Nebulae. A Study of the 
Emission Lines and Morphology.
\footnote{Based on observations made with the NASA/ESA Hubble Space Telescope,
obtained at the Space Telescope Science Institute, which is operated by the 
Association of Universities for Research in Astronomy, Inc., under NASA 
contract NAS 5--26555}}

\author{Letizia Stanghellini\altaffilmark{2}}
\affil{Space Telescope Science Institute, 3700 San Martin Drive,
Baltimore, Maryland 21218, USA; lstanghe@stsci.edu}

\author{Richard A. Shaw}
\affil{National Optical Astronomy Observatory, 950 N. Cherry Av.,
Tucson, AZ  85719, USA; shaw@noao.edu}

\author{Max Mutchler}
\affil{Space Telescope Science Institute, 3700 San Martin Drive,
Baltimore, Maryland 21218, USA; mutchler@stsci.edu}

\author{Stacy Palen, Bruce Balick}
\affil{Astronomy Department, Box 351580, University of Washington, Seattle 
WA 98195-1580; palen@astro.washington.edu, balick@astro.washington.edu}

\and

\author{J. Chris Blades}
\affil{Space Telescope Science Institute; blades@stsci.edu}

\altaffiltext{2}{Affiliated with the Astrophysics Division, Space Science
Department of ESA; on leave from INAF-Osservatorio Astronomico di Bologna}

\begin{abstract}

HST STIS slitless spectroscopy of LMC PNs is the ideal tool to study 
their morphology and their ionization structures at once. We present the
results from a group of 29 PNs that have been spatially
resolved, for the first time, in all the major optical lines. Images
in the light of \ha, \nii, and \oiii are presented, together with
line intensities, measured from 
the extracted 1D and 2D spectra. A study on the surface brightness in
the different optical lines, the electron densities, the ionized
masses, the excitation classes, and the extinction 
follows, illustrating an ideal
consistence with the previous results found by us on LMC PNs.
In particular, we find the surface brightness decline with the 
photometric radius to be the same in most emission lines. We find that
asymmetric PNs form a well defined cooling sequence in the excitation --
surface brightness plane, confirming their different origin, and 
larger progenitor mass.

\end{abstract}


\section {Introduction} 

Planetary Nebulae (PNs) have been used extensively to gain insight into
the late stages of evolution of low and intermediate mass stars. In
particular, the observed nebular morphology reflects the geometry of the
initial mass ejection during and shortly after the AGB phase of stellar
evolution followed by the cumulative dynamical effects of winds, heating
by ultraviolet photons, plus the sudden increase in pressure from the
passage of an ionization front.  It is now widely accepted that PNs are
the ejecta of $\sim1$ to 8 \sm~ progenitor stars, expelled at the end of
the thermally pulsating,  Mira--like phase on the AGB. These ejecta
travel at a low velocity, and  remove most of the stellar envelope
eventually exposing a white dwarf core.

Most PNs show some degree of asymmetry.  Classical round PNs, which
might represent isotropic mass loss, are the minority.  Balick (1987)
defined several morphological classes based on the outline of the nebular
core, such as elliptical and butterfly. 
Balick
argued that a large fraction of PNs have a relatively dense equatorial waistband.
Prolate elliptical, bipolar, and closely related
nebular geometries develop as a consequence.
\citet{man96a} use the
interior symmetry to define classes. The major morphological classes defined
by \citet{man00} are round (R), elliptical (E), 
bipolar or quadrupolar (B),
and pointsymmetric (P) planetary nebulae. Bipolar core (BC) PNs, defined
by \citet{sta99}, are 
those planetary nebulae characterized by a bi-nebulosity in the core,
wich may have lobes below detectability.

Bipolar PNs appear to be associated with particularly dense tori or other
types of collimation ``nozzles'' that form as a normal part of their
evolution \citep{bal87,ipb89}.
Interestingly, Galactic bipolar PNs are located preferentially in the
plane---i.e., their scale height (z) distribution is similar to that of
Pop I stars initially more massive than about 2 M$_\odot$.  Other
less extreme morphological types of PNs have a larger scale height and
evolve from less massive progenitors \citep{sta93,man00}.

Bipolar and other extremely axisymmetric PNs also
tend to be enriched in N and O, but relatively depleted in C/O as found,
e.g., by \citet{pei78,pei97}.  These results provide important insight
into the shortcomings of present models for AGB envelope ejection 
\citep{fra00}. \citet{pei78,pei97} suggested that convective dredging
operates differently in stars of different initial masses, as expected
from the models of stellar evolution \citep{van97}.

Therefore Galactic PNs suggest that initial stellar mass, and, perhaps,
local chemical composition determine the outcome of PN abundances and
morphologies. The trends found in the Galactic PNs are supported by a
sizable collection  of data; nonetheless, they suffer two key impediments
for this type of study: namely, the long-standing uncertainties in the
distances of Galactic PNs and the selection effects due to absorption by
interstellar dust.

One important aspect of morphological studies is the dependence of PN
morphology on the emission line in which they are imaged. It
has been demonstrated that, while generally the \ha and \oiii morphologies
nicely represent the overall nebular volume, low-excitation lines such
as \nii emphasize some particularly interesting
small-scale morphological features, such as organized ensembles of
low-ionization knots and point symmetry and
quadrupolarity \citep{sta93,man96b}.  A series of narrow-band images of PNs
is the ideal database to start answering some of the current
questions on the morphology formation mechanisms, and to study
morphological types in detail.

Statistical studies of the properties of PN morphologies, masses,
luminosities, ionization, and chemical abundances in which selection bias
is minimized and uncertainties in distances are eliminated can be pursued
in the Magellanic Clouds.  However, any comparison
of the morphological and stellar properties of Galactic and MC PNs require
the spatial resolution of the {\it HST} in order to resolve the nebulae
and to obtain accurate photometry of the central stars in very crowded
fields. Using {\it HST}, MC PNs can be observed with a physical resolution that is
comparable to ground-based observations of typical Galactic PNs.

Since 1999 we have embarked in a large study of MC PN morphologies,
evolution, central stars, and progenitors. At this time, several {\it
HST} programs that aim at collecting a very large dataset of STIS
slitless spectroscopy and broad band photometry of LMC and SMC PNs have
been completed or are active, and other programs
are scheduled to be executed in Cycle 10.  

This paper presents data from
the Cycle 8 {\it HST} snapshot survey of LMC PNs (program 8271),
consisting of a survey of 29 LMC PNs, acquired in broad band
imaging and slitless spectroscopy with STIS. The  broad band images of
these PNs have been already published by \citet{sha01} (hereafter Paper
I), with an extensive discussion on PN morphology and evolution for the
sample of program 8271 and of other HST archived images of MC PNs.
Here we present a study of the monochromatic images of the LMC PN
sample, obtained by using STIS slitless spectroscopy. 
After calibration, the slitless
images are functionally the same as calibrated narrow-band filter images
since Doppler smearing is insignificant.  We discuss the data acquisition
strategy and analysis procedures in $\S$2. In $\S$3 we present the
images, the integrated spectral line intensities, and surface brightness
estimates.  Some of the surface brightness results are
unexpectedly correlated, so we discuss the data trends and their
significance. Finally, in $\S$4, we give a summary of our work so far, and
a glance to the future of our extensive Magellanic Cloud Planetary
Nebulae project.

\section{Data calibration and analysis}

\subsection{Slitless strategy}

The observations presented in this paper are from {\it HST} GO program 8271, 
using the Space Telescope Imaging Spectrograph.  See Paper I for 
the observation log, observing configuration, target selection, acquisition, 
and basic calibration (through flat-fielding).  Paper I also presented, for 
each nebula, a broad-band image, a contour plot in the light of 
[\ion{O}{3}] $\lambda5007$, and a discussion of the morphological 
classification.  
We present here slitless spectrograms obtained with gratings 
G430M, covering the range 4818~\AA\ to 5104~\AA\ at 0.28~\AA\ pixel$^{-1}$, and 
G750M, covering the range 6295~\AA\ to 6867~\AA\ at 0.56~\AA\ pixel$^{-1}$.  See 
the STIS Instrument Handbook \citep{Leitherer_etal00} for additional details 
of the instrument setup. 
The exposures were planned to obtain a good signal-to-noise ratio in the 
[\ion{O}{3}] 5007 and H$\alpha$ emission lines, but up to eleven additional 
bright emission lines of varying ionization were detected, including 
H$\beta$ and [\ion{O}{3}] $\lambda$4959 using G430M, and 
[\ion{O}{1}] $\lambda\lambda$6300, 6363, 
[\ion{S}{3}] $\lambda$6312, 
[\ion{N}{2}] $\lambda\lambda$6548, 6584, 
[\ion{He}{1}] $\lambda$6678, and 
[\ion{S}{2}] $\lambda\lambda$6716, 6731 using G750M.  A stellar or nebular 
continuum was also detected in the spectrograms of some objects.

\subsection{One dimensional spectral extraction, and total line intensities}

For most nebulae the combination of dispersion and plate scale (0\farcs051 
pixel$^{-1}$) allows a clean separation in the dispersion direction of 
the monochromatic images for all emission lines. However, if the nebular 
extent exceeds about 1\farcs4 the H$\alpha$ and [\ion{N}{2}] $\lambda6548$ 
lines, and the [\ion{S}{2}] $\lambda6716$ and $\lambda6731$ lines, will 
overlap.  A few of the more extended targets suffered from this overlap, 
and a two-dimensional line deblending technique was required to solve for 
the individual narrow-band images (see \S~2.3).

We extracted one-dimensional (1D) spectra from the slitless 
spectrograms for each nebula and applied a photometric calibration using 
the standard STIS calibration pipeline module {\bf x1d} 
\citep{McGrath_etal99}. 

A significant complication when extracting extended objects is to 
include the majority of the light, while keeping the virtual extraction 
slit small so as to maximize the signal-to-noise (S/N) ratio 
\citep[see][]{Leitherer_Bohlin97}. 
The extraction box size for each nebula was chosen to accept most of the 
flux without compromising the S/N too much, and the background regions were 
carefully selected to avoid stray stellar spectra in the slitless images.  
In Table 1 we give the extraction parameters for the nebulae. 
Column (1) lists the LMC PN name (SMP nomenclature favored when available); 
columns (2) through (4) give respectively the grating name, the
center of the extraction 
window, and its size.

We measured emission line intensities using the IRAF\footnote{IRAF is distributed by the 
National Optical Astronomy 
Observatory, which is operated by the Association of Universities for 
Research in astronomy, Inc., under cooperative agreement with the National 
Science Foundation.} {\bf splot} task,
which permits flux measurements even when the intrinsic line profiles are 
far from instrumental (as they were for most of our targets).  
We typically 
use the deblending cursor command of {\bf splot}, even in cases where there 
is not any obvious blending of
emission lines. We prefer this procedure primarily because it automatically 
estimates the nebular continuum. We typically fit individual gaussian
widths for all the lines within a selected region, but we may fit a single gaussian width for multiple 
lines if we feel that the latter procedure 
achieves a better overall measurement. 

In cases where line profiles are significantly non-gaussian, and also for the
very extended objects, we may use the {\it e} cursor command in the {\bf splot} task.
We also use this command to check whether the total flux measured across 
several lines matches reasonably well with the
sum of the individual line fluxes measured with the deblending command. 
In a case with complex nebular continuum, we may
estimate the continuum better by eyeball, and measure it just
with cursor clicks, than by a fitting routine. 

In Table 2 we report the measured line intensities. Column (1) gives the
common names, Column (2) gives the
logarithmic \hb intensities, not corrected for extinction, in erg cm$^{-2}$ s$^{-1}$; column (3) 
lists the logarithmic optical 
extinction at \hb; columns (4) to (14) give the 
line intensities for each nebula, relative to \hb=100, 
not corrected for extinction. 
The line 
identifications, which are given in the column headings, were unambiguous 
in spite of the lack of a wavelength comparison arc since only the most 
prominent nebular lines are detectable in the spectrograms.

\subsection{Two-dimensional spectral extraction}

The geometric calibration of the 2D spectra is performed with the STIS
calibration pipeline module
{\bf x2d}, which produces a rectified 2D spectrum which is linear in both
the wavelength and the spatial directions. The output spectrum
has units of [ergs cm$^{-2}$ s$^{-1}$ \AA~$^{-1}$ arcsec$^{-2}$]. 
We used the 2D spectra to measure the 2D line intensity of all
detected spectral lines, using the same 
extraction windows and background offsets as
the 1D spectral extraction (see Tab. 1), so the two 
measurements are readily
comparable, and they offer a check on the procedure and on the calibrated
image quality. 

Two of the angularly largest of our targets, SMP~93 and SMP~59, suffered 
from severe overlap of the emission lines.  For these nebulae it was 
essential to apply a deblending technique on the 2D spectral images in 
order to separate the monochromatic images in the H$\alpha$-[\ion{N}{2}] 
group.  We used the following simple procedure: 
\begin{enumerate}
 \item {The images were sky-subtracted by using the mean signal within a
 suitable sky window (see Table 1 for the center and sizes of 
 the extraction windows).
 }
 \item {We take advantage of the fact that the [\ion{N}{2}] lines are 
 identical in shape and have a fixed intensity ratio.  This allows us to 
 use the redward, uncontaminated portion of $\lambda$6584 line to subtract 
 from the blend at the position of the $\lambda$6548 line; likewise the 
 uncontaminated blueward portion of the [\ion{N}{2}] $\lambda$6548 line 
 was used to subtract from the blend at the position of the $\lambda$6584
 line, leaving an uncontaminated H$\alpha$ image.
 }
 \item{We obtain the uncontaminated \nii image by subtracting the
 reconstructed \ha image from the entangled image group.}
 \end{enumerate}

A check on the quality of the above procedure has been performed by
comparing the total flux of the H$\alpha$ plus [\ion{N}{2}] blend 
with the sum of the flux from the blend as measured with the 1-D 
technique. The agreement is good at least to the 10$\%$ level in the PNs 
analyzed in this paper.  However, it is clear that the technique is far 
from perfect: the relative intensities for H$\alpha$ for both SMP~93 and 
SMP~59 is much less than what is possible for a nebular gas.  Our 
technique undoubtedly fails to assign the correct total flux in these line, 
and we report only the sum of the \ha and \nii intensities in Table 2. 
Furthermore, data relative to these two PNs will not be entered
in the diagnostic plots of this paper, wherever the extinction corrected
fluxes are used. 

Fow SMP~59 and SMP~93, the \sii $\lambda$6716-6731 lines are also
superimposed in the dispersion direction. In this case, since the line ratio
is not known (it depends on the electron density) we measure the 
line blend, and give it in Table 2.

\subsection{Error analysis}

The 1D and 2D intensities can be compared to make a detailed analysis of the
errors in our measurements.  The intensities for a given emission
line and nebula should be the same no matter the method used for the
analysis. On the other hand, we should limit this comparison to the nebulae 
whose images do not overlap in the dispersion direction.
We calculate the normalized difference between 1D and 2D measurements as 
$\Delta_{\rm 1D-2D}=(F_{\rm 1D}-F_{\rm 2D})/F_{\rm 1D}$ 
for each spectral line and each PN in our sample.
We plot $\Delta_{\rm 1D-2D}$ against the 1D flux in Figure 1.
We find that the difference 
depends strongly on the observed line intensity. In particular:
$|\Delta_{\rm 1D-2D}|<.05$ if ${\rm log} F>-12.25$,
$|\Delta_{\rm 1D-2D}|<.15$ if $-12.25<{\rm log} F<-12.75$, 
$|\Delta_{\rm 1D-2D}|<.2$ if $-12.75 < {\rm log} F<-13.5$, 
$|\Delta_{\rm 1D-2D}|<.25$ if $-13.5<{\rm log} F<-14.5$, and finally 
$|\Delta_{\rm 1D-2D}|<.55$ if ${\rm log} F>-14.5$. 
The only exception to these results are a handful of data points in the 
$-13.5<{\rm log} F<-14.5$ interval.
These are the [\ion{N}{2}] $\lambda$6584 lines that may be blended with the
H$\alpha$ lines, even for relatively small objects (e.~g., SMP~10 and SMP~100), and
another couple of lines whose measurements have very high errors (the [\ion{O}{1}] $\lambda$6363 line
in SMP~78 and 
SMP~9, and the He~I 
$\lambda$6678 line in SMP~81).

We have also examined the variance of $\Delta_{\rm 1D-2D}$ with respect to the nebular 
radii, to find that 
there is no dependence of the error in the flux measurements to the size 
of the nebula, nor on the morphological type.
We conclude that the errors listed above, in relation to the intensity
of the 1D spectral lines, are the formal internal observational errors for the 
fluxes presented in this paper.


In order to asses the quality of our data, and the improvement over the
published line intensities, we have compared the measured MC PN intensity 
ratios of Table 2 to the fluxes published by \citet{vdm}, 
\citet{dop91a}, and \citet{dop91b}. In all, we found published fluxes 
for 15 of  29 planetaries.  The line intensities from \citet{dop91a} 
and \citet{dop91b} are corrected for extinction, so it was necessary to 
apply the reddening function before comparing them to our data.  
To this end, we use the extinction constants given by the individual
references, and average Galactic reddening curve of \citet{Savage79},
which is
reliable at these wavelengths even for the LMC \citep{how}.
In Figure 2 we show the comparison between our line intensity ratios and
those from the references above. The intensity ratios are plotted in 
logarithmic form, to allow the simultaneous view of low and high fluxes. 
The correspondence of our intensities to the ones in the literature is 
generally good, and typically within the observing errors quoted in the 
references (indicated with vertical lines on the plot).
The notable exceptions are the [\ion{N}{2}] 6548 lines
of SMP~10 and SMP~100, overestimated in the references (the dots
in the middle of Fig. 2). We believe that our [\ion{N}{2}] 
measurements for these two nebulae are more reliable than those 
by the literature source \citep{vdm}. In fact, we know that the [\ion{N}{2}] 
6584/6548 ratio should be approximately 3 \citep{ost89}, and our ratios are
2.96 and 2.98 respectively for SMP~10 and SMP~100, while the ratios from
\citet{vdm} are 11.4 and 11.1 respectively. The errors in the reference may 
be caused by the poor separation of the [\ion{N}{2}] and H$\alpha$ lines in 
the dispersion direction. \citet{vdm}, for example, use a bandpass of FWHM 
16 \AA\ to observe the H$\alpha$ lines, and this separation may be not 
enough for SMP~10 and SMP~100, where the [\ion{N}{2}] lines have
low intensity.

A close inspection of Figure 2 reveals a small discrepancy in the [\ion{O}{3}] 
$\lambda\lambda4959, 5007$ line intensities (the clump of dots at log F $\approx$
2.5). The source of the mismatch may lie in a 
difference in the H$\beta$ fluxes, which are used to scale the intensities
and which are often weakly exposed in our spectrograms for larger nebulae. 
The ultimate cause of the discrepancy may lie with a systematic problem in 
the flux calibration for the STIS CCD.  
\citet{Stys_Walborn01} found that an apparent degradation in the sensitivity 
of the STIS CCD with time (which is not yet corrected in the calibration 
pipeline) may in fact have its root cause in a degradation of the CCD 
charge transfer efficiency.  While not yet well characterized for the 
grating/central wavelength combinations used here, in general the effect 
is that charge lost during the CCD readout is manifested as lower measured 
fluxes.  Depending upon the brightness of the surrounding mean sky and the 
number of detected counts per pixel of the emission feature, the effect 
could range from less than a percent to as much as 15\% \citep{Stys_Walborn01} 
for the emission lines reported here.  We anticipate that the effect for the 
H$\beta$ fluxes (and hence the scaling for all of the emission lines) for 
angularly large nebulae is likely in the range of 1--3\%.  The effect on 
the very weakest lines (such as [\ion{O}{1}]) in large nebulae may approach 
10\%, but the effect on well exposed lines in small nebulae is likely 
negligible.

\section{Results}

\subsection{Nebular morphology in various lines }

The calibrated slitless images of all but the very largest LMC
PNs are shown in Figure 3.  For each nebula we show the \oiii 5007~\AA\
section of the G430M spectra, when available, and 
the \ha and \nii line group section of
the G750M spectra.  The images of SMP~94 are not shown in Figure 3,
because we are convinced this object is not a PN, based on its
spectral structure (see Paper I).
The slitless images for the two largest nebulae are shown in Figure 4,
where we show the \oiii 5007 \AA~ images,  the \ha and \nii group, and
the \nii images as reconstructed with the algorithm described in $\S$2.3.

Paper I describes the morphology of our sample nebulae in detail. In the
following we limit the description to interesting morphological
features of the spectra that have not been mentioned in Paper I.
Morphological types are taken from Paper I unless stated otherwise.
As in the other papers of this series, we use the \citet{man96a} 
morphological classification scheme, and we also add the {\it bipolar core} 
class. 
Hereafter, when we refer to {\bf symmetric} PNs we include E and R {\it without} a 
detected bipolar core, and with
{\bf asymmetric} PNs we indicate the bipolar and quadrupolar (B), and
the bipolar core (Rbc and Ebc) PNs.
We keep pointsymmetric PNs outside these groups. 

\begin{itemize}

\item{J~41:  The nebular morphology is Ebc. It is observed
only with the G750M configuration, and only the \ha emission is
detected.  No low-ionization structures (hereafter ``LIS''') are seen in
this object.}

\item{SMP~4:  The nebular morphology is E. This attached
halo multiple shell PN shows its morphology in all the  detected lines.
This is shown for \oiii $\lambda$5007~\AA\ and \ha in Figure 3, and it is
also true for \hb and \oiii $\lambda$5007~\AA.  The emission in the \nii
lines was marginally detected below the 3$\sigma$ level. No LIS' are
seen in this object.}

\item{SMP~9:  The nebular morphology is Ebc. The barrel shape
is shown in all detected emission lines, including \nii.   No LIS' are
seen.}

\item{SMP~10: The nebular morphology is P. The \nii emission occurs
only in the {\it arms} of this PN. There are three
distinct morphological features: an inner ellipse (not visible in \nii), a
round edge-brightened mantle that surrounds it, and two ``spiral arms''
that are most prominent in \nii.  We note that the enhanced emission that 
defines the inner ellipse 
is oriented such that the major axis is aligned with the {\it arm} structure 
in the outer nebula. It is conceivable that this feature traces more subtle, 
unresolved structure such as a jet from the central star that terminates 
at the ends of the arms, similar to the Cat's Eye nebula (NGC 6543). 
Note that the photometric radius
measured by \citet{sha01} on the \oiii image
is very close to a photometric  radius measured
on the \nii image. Although this nebula is rather extended, none of 2D
the spectral lines overlap.}

\item{SMP~13: The nebular morphology is Rbc. The same patchy morphology
is visible in all detected lines. The \nii emission (see Fig. 3) is very
faint except along the lobe edges where its structure is clumpy. This
behavior is characteristic of Galactic bipolar PNs such as NGC~2346 and
NGC~6537, which are similar in appearance to SMP~13. In the spectral data
we can see a faint stellar spectrum that could be associated with the
central star.}

\item{SMP~16: The nebular morphology is B. This very prominent bipolar
shows its morphology in all detected lines.  As mentioned in Paper I, the
\nii emission reveals a knotty character of the lobes,  marginally
evident in Fig. 3. No LIS' are seen in this object.}

\item{SMP~18:  The nebular morphology is Rbc. A central cavity of this
multiple shell PN may not be evident in  Fig. 3, but it is obvious in the
frame. The faint stellar spectrum seen in  Fig. 3 may be associated with
the central star.}

\item{SMP~19:  The nebular morphology is Ebc. The nebula reminds us of a
barrel seen in projection.  Small protrusions may emerge from the barrel
along its symmetry axis, much like NGC~40.  No LIS' are seen.}

\item{SMP~25: The nebular morphology is R. The nebula appears amorphous
in this logarithmic display.  No LIS' are seen.}

\item{SMP~27: The nebular morphology is Q. This interesting
quadrupolar PN has the central star spectrum easily seen both in the
G430M and G750M observations.  The \nii emission has very low surface
brightness.}

\item{SMP~28: The nebular morphology is P. The {\it arms} seen in the
\nii image are too thin to be reproduced in Figure 3. The shape may be
similar to that of SMP~10.}

\item{SMP~30: The morphological type is irregular, possibly B.
The characteristic of this PN is the very high \nii intensity
at the nebular edges. The \nii images give much more information on the
nebular morphology than the \ha or \oiii images, defining knots and arcs.}

\item{SMP~31: The nebular morphology is possibly R. A central star is not seen
in the clear image, but there is hint of the  stellar continuum in the
slitless spectra.  No LIS' are seen.}

\item{SMP~34: The nebular morphology is E. A different morphology is
detected in the \nii lines. Although these lines are faint, a hint of a
bipolar shape is shown.  No LIS' are seen.}

\item{SMP~46: The nebular morphology is Ebc. A central cavity is
particularly clear in \nii.  No LIS' are seen.}

\item{SMP~53: The nebular morphology is E, possibly Ebc. A ringlike structure is
clearly seen in the \nii emission lines.  Filamentary LIS' surround the
core, something like SMP~30.}

\item{SMP~58:  The nebular morphology may be R. A barely resolved PN, the
central star spectrum is seen in the  G750M slitless observations (see
Fig. 3). An emission feature at about 6577\AA\ is probably stellar. No
LIS' are seen.}

\item{SMP~59: The nebular morphology is probably Q. A spectacular 
PN, its morphology is particularly evident in the \nii emission lines.
The reconstructed \nii image in Fig. 4 shows that there may be multiple
emission or knots in this nebula.}

\item{SMP~65: The nebular morphology is R. No LIS' are seen.}

\item{SMP~71: The nebular morphology is E. No LIS' are seen.}

\item{SMP~78: The nebular morphology is Ebc. No LIS' are seen.}

\item{SMP~79: The nebular morphology is Ebc. Ringlike structures are
evident in the \nii emission.  No LIS' are seen.}

\item{SMP~80: The nebular morphology is R. Ringlike structures are
evident in the \nii emission, and the central star spectrum is seen in
the slitless observations with the G750M grism.}

\item{SMP~81 is not shown in Figure 3 for lack of interesting extended
features. The nebular morphology is probably R. An extremely faint, marginally
detected central star spectrum is seen in the G750M slitless
observations.}

\item{SMP~93: The nebular morphology is B. The best way to study the
morphology of this spectacular bipolar PN is to look at the \nii
reconstructed image of Fig. 4. The central ring and the lobes are
enhanced at this wavelength. The morphology of the \ha emission looks
similar to that of the \oiii emission. The edges of the main
morphological structures are low in emission, much like most Galactic
PNs.  NGC~6445 is similar in structure to SMP~93.}

\item{SMP~94: The nebular morphology is unresolved. We believe that this object
is not a PN (see the discussion in Paper I).
Although we include the line flux analysis in Table 2 for
SMP~94, we do not include it in the Figures, not in the
following  analysis and related
plots.}

\item{SMP~95: The nebular morphology is Ebc. Faint low-ionization ansae
can be seen in \nii.}

\item{SMP~100: The nebular morphology is Ebc, or Q. The \nii emission lines
show the knots' outline of this interesting PN. No LIS' are seen.}

\item{SMP~102: The nebular morphology is R, possibly bc. The interior shows a round
clumpy ring. The brightest of the knots is faintly visible in \nii.}

\end{itemize}

 From the above analysis we note that, while the morphological type is
more or less the same in all detected lines for most PN, the
morphologies are sometimes more evident in the low-ionization images,
as if these arise primarily along the nebular border. Some distinct,
compact emission features are seen most prominently in \nii.

\subsection{The surface brightness-radius correlation}

In Paper I we found a tight correlation between \oiii$\lambda$5007~\AA\
surface brightness (hereafter SB, defined to be the integrated 
line intensity divided by the nebular area $\pi {R_{\rm phot}}^2$) 
and nebular photometric radius $R_{\rm phot}$
which appears to be independent of morphological type or other
properties of the targets. Here we extend
these studies to other bright emission lines in order to see how the
correlation depends on the nebular ionization state, abundances, or other
parameters. In the discussion below we consider the \oiii and \siii
lines to arise in the nebular zone of moderate ionization, \nii and \sii
to arise in zones of low ionization, \ha and \hb to arise in both zones,
and \oi to arise in an excited neutral zone at the nebular boundary.

In Figure 5 we plot log SB against log $R_{\rm phot}$ (the physical photometric 
radii) for each of the
representative emission lines of our nebulae
(except \sii which are presented in Figure 6 and discussed later). 
All surface brightnesses
have been corrected for extinction using the extinction constants of
Table 2 and the Galactic extinction curve (see $\S$~2.3). 
Note the
consistency in the trends from one line to another, with the exception of
\nii, where the overall relation is
hidden by scatter. The reason for the scatter is that the bulk of the 
\nii emission
often arises in the nebular interior -- in the same zone -- where the
\oiii lines arise, whereas in other cases the \nii lines arise
at the nebular boundary, or in the LIS'. This means that the plot 
of the surface brightness decline in the \nii line is related not
only to the evolution of the gas, but also to the ionization effects.
Note that
the most discrepant points are relative to asymmetric PNs.
The agreement of the \oi line
SB vs. $R_{\rm phot}$ correlation with \oiii and \ha plots is the big
surprise: \oi should arise only at the nebular boundary, and still it 
declines with R in a similar fashion as the \oiii lines.

In Figure 5 we superimpose lines describing the 
SB $\propto {R_{\rm phot}}^{-3}$ relation, the best eye-fit to 
the data for all emission line plots.
Once
again we see that the trends are tight, and the trends seem to be
independent of the nebular state or morphology. By examining the
SB--$R_{\rm phot}$ trends for each morphology type
we see that while the SB trend is the same for all morphological types, 
asymmetric PNs are located in the bottom right part of the relation, 
indicating their evolved status. A similar result, but for the \oiii
line only, was derived in Paper I\footnote{Note that the upper left panel
of Figure 5 is not identical to the log SB--log $R_{\rm phot}$
Figure of Paper I, for two reasons. First, in Paper I we added 
in the Figure all data from the archived LMC PNs as well, while
here we restrict the discussion to the data of our program 8271. Second, 
in Figure 5 we updated all the line intensities, and relative
surface brightness, by using our own HST fluxes from Table 2.}

The relations above hold
only in the cases in which the nebular density $N_{\rm e}$ is smaller
than the  critical density, $N_{\rm crit}$ (the density at which the
collisional de-excitation rate balances the radiative  transition rate).
We have calculated the critical density for the ions  considered in
Figure 5, by using the {\bf nebular} routines \citep{sha95}, to prove
that $N_{\rm e} \le N_{\rm crit}$ for all the PNs for the ions examined
in the Figure.

Let us now examine the cases in which the electron density of the PNs is
higher  than the critical density. In Figure 6 we plot the
usual log SB vs. log$R_{\rm phot}$ relation,  in the light of \sii.  The
critical densities are respectively $N_{\rm
crit} \sim1 3.6 \times 10^3$ cm$^{-3}$ for the transition from  level 2
to level 1 (6731 \AA~ line), and $N_{\rm crit} \sim1 1.4 \times 10^3$
cm$^{-3}$ for the transition from level 3 to level 1 (6716 \AA~ line).
We have compared these critical densities with the electron  density of
the PNs, and plotted  with filled symbols in Figure 6 (we do not 
code for morphological type in this Figure, to avoid confusion) the PNs in
which  $N_{\rm e} \ge N_{\rm crit}$. We can infer
that a simple surface brightness versus photometric radius relation does
not hold for the majority  of the objects, in this line.

On the other hand, most nebulae are less dense than the critical density
for the \sii$\lambda$6716\AA\ line, and a sizable fraction of them are
less dense than the critical density for the \sii$\lambda$6731\AA\ line.
This  effect explains the different relation observed.

\subsection{Electron density and ionized masses}

Nebular densities $N_{\rm e}$ of the 12 MC PNs with the brightest \sii
lines were estimated using the IRAF package {\bf nebular} and assuming an
electron temperature of 10$^4$ K.  The electron densities calculated in
this way are reliable only for the density interval between ~300 and 7000
cm$^{-3}$ \citep{sk89}.  The log N$_{\rm e}$ -- log R$_{\rm phot}$ is shown in Figure
7. Morphological types are plotted with different symbols.  Based on a
small and selected sample, the densities of MC PNs are not related to their
morphologies.

If all PNs have the same mass, morphology, expansion rate, and density
boundedness (i.e., ultraviolet opacity) then we would expect a tight
negative correlation between density and radius.  We see only a loose
correlation, if any at all.  This is no surprise. The same absence of a
tight density--radius correlation was seen earlier by 
\citet{dop90} based on ground-based data.

The ionized masses have been calculated from the \sii densities and
the photometric radii using the methodology of \citet{bs}. This method
assumes spherical geometry and uniform density within the filled 
parts of the nebula, which is obviously the
exception and not the rule (Figs. 3 and 4), but does not assume 
a value for the filling factor ($\epsilon$) of the nebulae \citep{bs}.
Implicit also is the
assumption that the \sii densities are characteristic of the entire
nebula, whereas we have seen that knotty and filamentary LIS' sometimes
dominate the \nii and \sii emission.

We find that the average ionized mass is 0.21 \sm, the same that
has been found by \citet{bs} from a larger LMC PN sample, albeit without
diameters measured directly from deep {\it HST} images. The ionized
masses for the only bipolar and elliptical PNs in our sample with reliable electron
density are respectively 0.33 \sm and 0.17 \sm.
\citet{bar87} used the \oii electron density to derive the ionized mass of
32 Magellanic Cloud PNs, and they found an average mass of 0.25 $\pm$ 0.17 \sm
for the complete sample, and 0.39 $\pm$ 0.22 \sm for the type I PNs
within their sample. Our results are thus in agreement with those of
\citet{bar87}, including a possible ionized mass offset between
PN types, type I PNs and bipolar PNs being the more massive ones.
A much larger data set is needed to confirm these trends.

The ionized mass
versus radius logarithmic relation is plotted in Figure 8 using the same
symbol definitions as in Figure 7.  The tight correlation found for
Galactic PNs \citep{bs} is not found in this plot of LMC PNs, where the two
(logarithmic) variables have a low correlation coefficient of 0.35.
Much of this scatter is the result of inappropriate assumptions such as
spherical symmetry, constant density, and accurate nebular radii.  The
assignment of a nebular radius is especially treacherous for bipolar PNs
with open, slowly fading lobes along the major axis and bright, sharp and
closely spaced edges along the minor axis. 

\subsection{Nebular excitation}

In planetary nebulae,
one measure of stellar evolutionary state is
nebular excitation, which tracks the evolution
of the 
temperature of the central star.  For ionization bounded nebulae this can
be gauged from the relative volumes of regions dominated by He$^{\circ}$,
He$^{\rm +}$ and He$^{\rm ++}$. Nebular excitation is best measured using
ratios of recombination lines of H and He which are only slightly affected
by electron temperature T$_{\rm e}$.  The signal-to-noise and restricted
bandwidths of our spectra render this impractical.

We use an alternate measure of excitation, EC=(I$_{\rm [O~III]5007}$
+I$_{\rm 4959}$)/I$_{\rm H{\beta}}$. The ratio is sensitive 
linearly to the fraction of O$^{\rm ++}$/O (much like He$^{\rm +}$/He) and
exponentially to the electron temperature T$_{\rm e}$.  The fractional
ionization of O$^{\rm ++}$ varies as the star evolves: over time the
dominant stages of oxygen ionization first rises very quickly from
O$^{\rm +}$ and then very slowly to O$^{\rm +++}$.
Similarly
T$_{\rm e}$ tends to increase with time as the average kinetic energy of
photoionized electrons rises with stellar surface temperature.  As the
central star evolves from 25,000 K to over 10$^5$ K, EC will rise. Afterwards, 
when the star begins to fade and lower its temperature, the EC will
begin to fall.  Overall, however, EC is a decent estimator of stellar temperature.

The measured photometric radius,
R$_{\rm phot}$ should be a good measure of age because the nebula
expands.  But it is far from perfect.  Different rates of expansion in
different PNs will introduce scatter into the use of R$_{\rm phot}$ as an
age indicator for a group of PNs (see Paper I).

Since round and elliptical PNs seem to evolve from progenitors of
lower mass than bipolar and bipolar core PNs (see Stanghellini et al. 2000)
we
might expect to find correlations in EC and R$_{\rm phot}$ separately for
various morphological types.  Figure 9 is a plot of EC vs R$_{\rm phot}$ for
LMC PNs. It shows a good correlation for asymmetric PNs (the squares and
triangles,
see the caption) and scatter for other
types.  We find small low-excitation objects and relatively large
high-excitation R and E PNs, as expected.  The scatter associated with
these objects could be related to the different rates at which they
expand as well as a bias against the discovery of large, low-SB PNs.

Figure 10 is a plot of EC vs log SB \oiii for LMC PNs.  The interpretation of the
plot is complex and limited by the small sample size. Assuming that
all asymmetric (squares and triangles) PNs were formed with a similar 
O/H ratio it follows that they show a clear correlation independent of their
initial oxygen abundances.  
On the other hand, no correlation is seen for R and E PNs.
There is a bias against the discovery of large R and E PNs owing to their
very low surface brightnesses. But most of all we believe that the
fraction of escaping ultraviolet radiation increases drastically for R
and E PNs. Variations in mass or expansion speed will tend to introduce
scatter into SB for a group of PNs.

The \ha recombination-line luminosity of an expanding PN is a measure of
the rate at which ionizing photons are being absorbed in the nebula.  We
expect the \ha flux to be constant while the nebula is ionization bounded
and to drop thereafter until the star itself drops in UV photon emission
rates late in its life. In Figure 11 we plot the \ha line luminosity, corrected
for extinction, against EC. Since all PNs in this plot are at the same
distance, the \ha intensity is proportional to the stellar luminosity. 
Note that in Figure 11 the excitation constant 
increases from right to
left in the plot, as does the effective temperature in the HR diagram.

\citet{dop90} have used a similar diagram to compare the observed LMC and
SMC PNs  with simple evolutionary models. The novelty of our approach
is that we added the information on nebular morphology.  In Figure 11 we
can follow the PN evolution in {\it two dimensions}, exactly as we follow
the stellar evolution in the HR diagram. We can see the simultaneous
effects of changes in the effective temperature (through the parameter
EC) and of the luminosity (through the \ha brightness) for different
nebular morphologies. PNs evolve first from right to left (and from
bottom to top) as their central stars get hotter and their shells get
thinner, then from top left to bottom right as the central stars  evolve
in the cooling line, at constant radii.

Naturally, much modeling is needed before we could derive a quantitative
conclusion from Figure 11, but we feel that the empirical result is
very strong, that is, most of the observations of asymmetric PNs in the
LMC  are consistent with them being on the {\it cooling sequence}, beyond the
central star's temperature maxima. The opposite is true for symmetric
PNs.  This observational result is consistent with our other findings
that relate asymmetry with higher progenitor mass, in this case it looks
like also the post-AGB stellar mass is higher for asymmetric PNs.
Our future study on the central star's photometry will clarify some of the
open questions posed by these diagrams, and appropriate hydrodynamic
modeling will complete the description of the evolution of PN in the LMC.

\subsection {Nebular extinction}

We analyzed the extinction in this sample of nebulae to look for evidence 
of internal dust, and for any variations with morphological type. 
\citet{Ciardullo_Jacoby99} reported evidence of a correlation of extinction 
with the mass of the central star for PNs in the Magellanic Clouds and in 
M31.  We examined some nebulae in our sample to look for spatial variations 
in the extinction, which we believe would be a strong indicator of the 
presence of dust internal to the nebulae.  We took ratios of the H$\beta$ 
and H$\alpha$ images for PNs where the extinction constant was relatively 
high, or that had morphological features that can be associated with the 
presence of internal dust, such as the pinched waist of the bipolar nebulae 
SMP~16 and SMP~30.  In no case did we find significant spatial variations 
in H$\beta$/H$\alpha$, at least on spatial scales of $\sim0.04$ pc.  We note, 
however, that many of the PNs with high extinction are angularly small, and 
furthermore the S/N ratio of our H$\beta$ images is often not high.  Still, 
for nebulae within our sample that lend themselves to this analysis, the 
internal extinction, where present, must be spatially uniform on small
spatial scales.

Another way to address the question of whether the extinction is internal 
to the nebulae is to examine whether it declines with nebular physical 
radius, as might be expected if internal dust were either destroyed or 
geometrically diluted as the nebula expands.  
We plot in Figure 12 the nebular extinction 
constant vs. the log of the physical radius for all nebulae in this sample,
and we also include the PNs in the archived sample by Dopita (HST
program 6407, see also Paper I).
The Figure shows that the amount of extinction does not depend 
on morphological type, except possibly for large bipolar nebulae where $c$ 
is generally low.  It also shows that the amount of extinction depends only 
weakly on physical size, in that nebulae smaller than 0.1 pc suffer at 
least some extinction.  This question should be examined again with a 
larger data set, but for now the safest conclusion is that the extinction 
for PNs in this sample spans a large range at all sizes and does not depend 
strongly, if at all, on morphological type.  \citet{sta00} showed 
for Large Magellanic Cloud PNs that bipolar (and bipolar-core) morphology 
correlates with chemical properties that are related to higher-mass central 
stars.  Yet the absence here of any relation between bipolar morphology 
and extinction, and the lack of internal variation in extinction on modest 
spatial scales, suggests that there is no clear relationship between 
internal nebular extinction and central star mass.  The lack of nebulae 
smaller than 0.1 pc with zero extinction may indicate the presence of dust 
within these nebulae; our spatial resolution is insufficient to search for 
spatial variations in extinction.  It would be worthwhile to observe these 
nebulae in the thermal infrared to examine the dust properties more 
thoroughly.  In any case, we find no evidence that internal extinction 
in LMC PNs varies with central star core mass.

\section{Discussion}

In this paper we have shown the STIS slitless observation of a group of
29 LMC PNs, and analyzed the spectroscopic data. The images of Figures 3
and 4 present an  unique data set that includes a wealth of morphological
and spectral information. The line intensities measured from the slitless
spectra, although limited to the brightest  spectral lines in the
observed wavelength windows, represent some of  the best quality spectral
data for the LMC PNs. 
By adding spatial information to the spectrograms, we are able to 
determine an accurate morphology, where the effects of the ionization 
structure are fully understood. Our comparison of the 1D to the 2D line 
fluxes helps us to characterize our internal consistency, while the 
comparison to published data reveals the high photometric quality of 
these data

The nebular emission at the different wavelengths has been analyzed in
various ways. We obtain the following results.

\begin{enumerate}

\item{While the \ha and \oiii lines are perfect for morphological studies
of the main body of the PNs, the \nii emission lines are ideal to detect
knots, ansae, and other deviation from symmetry, not always detectable in
the \ha light, and to clarify the morphological classification.}

\item{The surface brightness relation with the photometric nebular radius
is indistinguishible for all the observed lines where the nebular electron
density is smaller than the critical density, and with the exception of
the \nii lines.  The very tight, universal log SB $\propto$ -3 
R$_{\rm phot}$ relation indicates the nebular evolution.
Such a relation could be used to calibrate the Galactic PN distance scale.}

\item{The marginal correlations between the photometric radii and the
electron  densities and the ionized nebular masses indicate
that the spherical approximation is wrong for the LMC PNs, and that
assuming that all PNs have the same ionized mass is very misleading. We
found that the average ionized mass for  our sample is 0.21 \sm. A
marginal correlation between asymmetry and high ionized mass was
detected.}

\item{The \oiii to \hb line ratio is a good measure of the nebular
excitation. We find that there is a tight correlation 
between EC and size in asymmetric PNs, suggesting that these nebulae
are on a faster evolving track. 
This result is in agreement with our
previous results of Paper I, and \citet{sta00}, that observations of
asymmetric LMC PNs are in accord with a rapidly evolving, more massive
progenitor, younger generation planetary nebula sequence, compared to
the low mass, slow evolving, round PNs.}

\item{The internal extinction of the observed nebulae is generally low,
and does not depend markedly on morphological type. We find no evidence for a
relation between the progenitor mass indicators and the extinction
of the nebulae}.

\end{enumerate}
In the future, we plan to extend this analysis to our other data samples,
the SMC PNs that have been observed in Cycle 9, and an expanded
LMC data
set that is being collected at the time of writing. At the same time, for
the dataset presented in this paper, we will also extract the information
about the central stars of those PNs. Eventually we will relate the
stellar and nebular properties of all the samples in study.

While most of the results presented in this paper (especially from points
2 and 4 above) need extensive nebular modeling to be understood, their
empirical value is nonetheless remarkable. We confirm, with this work,
the importance of slitless spectroscopy for maximizing the rate of useful
information of these objects. Our future plans include hydrodynamical
modeling of the SB vs. R$_{\rm phot}$ relation to test its physical
meaning, and to asses its validity as a tool to build the LMC based PN
distance scale.  We have much to understand about the shape and
universality of this relationship which may be useful for PN
distances as the period-luminosity relationship for Cepheid variable stars.

Our Web page,
http://archive.stsci.edu/hst/mcpn/,
includes many of the published data and data analysis, and it is a part of
the Hubble Data Archive.

This work was supported by NASA through grant GO-08271.01-97A from Space
Telescope Science Institute, which is operated by the Association of 
Universities for Research in Astronomy, Incorporated, under NASA contract
NAS -26555. We thank an anonymous referee for an essential suggestion.

\clearpage

\figcaption{Comparison of the 1D to the 2D 
line intensity measurement
from our {\it HST} data. We plot $\Delta_{\rm 1D - 2D}$ against the 
logarithmic 1D flux
for all nebulae of our sample and for all available lines. We do not plot data
for those nebulae whose lines
overlap in the dispersion direction, see text.}

\figcaption{Comparison of the measured line intensity ratios
with those in the literature, in logarithmic scale.
The vertical lines represent the
minimum errorbars for the reference fluxes.}

\figcaption{\oiii,\ha, and \nii images of the program nebulae 
(all PNs except SMP~59 and SMP~93).}

\figcaption{\oiii,\ha, and \nii images of SMP~59 and SMP~93.}

\figcaption{Surface brightness decline for the mutliwavelength images of the 
PNs in our STIS survey. Emission lines for which the SB is derived 
are indicated in the panels. Symbols 
indicate morphological types: open circles= round, filled circles=
pointsymmetric, stars= elliptical, filled triangles=bipolar core, filled
squares= bipolar (and quadrupolar) planetary nebulae.
The photometric radii are measured from the \oiii$\lambda$5007 \AA~ 
images (see text).}

\figcaption{
The log SB vs. log R$_{\rm phot}$ relation for the \sii  lines. 
Filled symbols:  PNs with $N_{\rm e} \ge N_{\rm crit}$, 
for \sii 6716 \AA~ (left panel) and \sii 6731
\AA~ (right panel). Points are not coded for morphology, to avoid confusion.}

\figcaption{Logarithmic electron density from the \sii doublet vs. log R$_{\rm phot}$. }

\figcaption{Logarithmic ionized mass vs. log R$_{\rm phot}$.}

\figcaption{Excitation constant, as defined in the text, plotted
against the logarithmic photometric radius.}

\figcaption{Logarithmic \oiii surface brightness vs. excitation constant.}

\figcaption{Logarithmic \ha line intensity vs. excitation constant. }

\figcaption{
Logarithmic extinction at H$\beta$ $\lambda4861$ vs. log R$_{\rm phot}$. 
}


\clearpage

%

%

\begin{deluxetable}{llcc}
\tabletypesize{\scriptsize}
\tablewidth{0pt}
\tablecaption {Spectrum Extraction Parameters \label{Extract}}
\tablehead {
\colhead {} & \colhead {} & \colhead {Ap.~Center} & \colhead {Ap.~Size}  \\
\colhead {Nebula} & \colhead {Grating} & \colhead {(pix)} & \colhead {(pix)}  \\
\colhead {(1)} & \colhead {(2)} & \colhead {(3)} & \colhead {(4)} 
}
\startdata  
J 41    & G750M & 517 & 23  \\
SMP 4   & G430M & 542 & 30  \\
        & G750M & 537 & 30  \\
SMP 9   & G430M & 521 & 20  \\
        & G750M & 517 & 20  \\
SMP 10  & G430M & 526 & 55  \\
        & G750M & 527 & 55  \\
SMP 13  & G430M & 531 & 27  \\
        & G750M & 526 & 27  \\
SMP 16  & G430M & 517 & 53  \\
        & G750M & 514 & 53 \\
SMP 18  & G430M & 553 & 27  \\
        & G750M & 548 & 27  \\
SMP 19  & G430M & 521 & 25  \\
        & G750M & 517 & 25  \\
SMP 25  & G430M & 511 & 15  \\
        & G750M & 506 & 15 \\
SMP 27  & G430M & 539 & 23  \\
        & G750M & 535 & 23  \\
SMP 28  & G430M & 503 & 12  \\
        & G750M & 508 & 23  \\
SMP 30  & G430M & 528 & 51  \\
        & G750M & 525 & 51  \\
SMP 31  & G430M & 513 & 23  \\
        & G750M & 508 & 11  \\
SMP 34  & G430M & 518 & 23 \\
        & G750M & 514 & 20  \\
SMP 46  & G430M & 538 & 23  \\
        & G750M & 533 & 23  \\
SMP 53  & G430M & 521 & 25  \\
        & G750M & 516 & 25  \\
SMP 58  & G430M & 521 & 23  \\
        & G750M & 516 & 23  \\
SMP 59\tablenotemark{a} & G430M & 521 & 25 \\
        & G750M & 516 & 25 \\
SMP 65  & G430M & 521 & 25  \\
        & G750M & 516 & 25  \\
SMP 71  & G430M & 521 & 23  \\
        & G750M & 516 & 23  \\
SMP 78  & G430M & 521 & 23  \\
        & G750M & 516 & 23  \\
SMP 79  & G430M & 521 & 23  \\
        & G750M & 516 & 23  \\
SMP 80  & G430M & 521 & 15  \\
        & G750M & 516 & 15 \\
SMP 81  & G430M & 521 & 23  \\
        & G750M & 516 & 11  \\
SMP 93\tablenotemark{a} & G430M & 521 & 25  \\
        & G750M & 516 & 25  \\
SMP 94  & G430M & 521 & 23  \\
        & G750M & 516 & 11  \\
SMP 95  & G430M & 521 & 27  \\
        & G750M & 516 & 27  \\
SMP 100 & G430M & 521 & 43  \\
        & G750M & 516 & 45 \\
SMP 102 & G430M & 521 & 29 \\
        & G750M & 516 & 29 \\
\enddata  
\tablenotetext{a}{Large angular size of nebula requires 2-D extraction; see text.}
\end{deluxetable}

%

\begin{deluxetable}{lrrrrrrrrrrrrr}
\tabletypesize{\scriptsize}
\rotate  
\tablewidth{0pt}
\tablecaption {Relative Emission Line Intensities of LMC Planetary Nebulae \label{Flux}}
\tablehead {
\colhead {} & \colhead {$F$(H$\beta$)} & \colhead {} & \colhead {\oiii} & 
\colhead {\oiii} & \colhead {\oi} & \colhead {\siii} &
\colhead {\oi} & \colhead {\nii} & \colhead {H$\alpha$} &
\colhead {\nii} & \colhead {\hei} & \colhead {\sii} &
\colhead {\sii} \\
\colhead {Nebula} & \colhead {(4861)} & \colhead {$c$} & \colhead {(4959)} & 
\colhead {(5007)} & \colhead {(6300)} & \colhead {6312} & \colhead {6363} & 
\colhead {(6548)} & \colhead {(6563)} & \colhead {6584} & \colhead {6678} & 
\colhead {(6716)} & \colhead {(6731)} \\
\colhead {(1)} & \colhead {(2)} & \colhead {(3)} & \colhead {(4)} & 
\colhead {(5)} & \colhead {(6)} & \colhead {(7)} & \colhead {(8)} & 
\colhead {(9)} & \colhead {(10)} & \colhead {(11)} & \colhead {(12)} & 
\colhead {(13)} & \colhead {(14)}  
}
\startdata  
J 41     &   & 0.00 &   &   &      &       &       &      &   &       &      &      &      \\
SMP 4   & $-13.55$ & 0.12 & 452.6 & 1354.4 & \nodata & \nodata & \nodata & \nodata & 312.6 & \nodata & \nodata & \nodata & \nodata \\
SMP 9   & $-13.43$ & 0.22 & 313.7 &  933.0 &    38.3 &     6.5 &    11.0 &    62.2 & 340.5 &   191.4 &     3.9 &    24.0 & 31.4 \\
SMP 10  & $-13.12$ & 0.16 & 377.2 & 1115.5 & \nodata & \nodata & \nodata &     6.1 & 324.0 &    19.0 & \nodata &     0.0 & \nodata \\
SMP 13  & $-12.88$ & 0.09 & 353.8 & 1068.2 &     4.2 &     1.4 &     1.5 &     7.2 & 306.1 &    15.0 &     2.9 &     3.1 &  3.3 \\
SMP 16  & $-13.22$ & 0.14 & 293.7 &  843.2 &    26.6 & \nodata &     9.3 &   196.4 & 318.5 &   628.7 & \nodata &    42.4 & 44.9 \\
SMP 18  & $-13.46$ & 0.07 & 274.0 &  841.0 & \nodata & \nodata & \nodata & \nodata & 300.6 & \nodata &     3.6 & \nodata & \nodata \\
SMP 19  & $-12.88$ & 0.18 & 442.7 & 1320.6 &    16.0 &     3.9 &     5.9 &    36.8 & 329.8 &   109.9 &     3.5 &    13.7 & 20.8 \\
SMP 25  & $-12.39$ & 0.12 & 268.3 &  800.0 &     3.8 &     1.1 &     1.3 &     7.2 & 314.6 &    22.7 &     4.5 &     1.1 &  2.1 \\
SMP 27  & $-13.54$ & 0.06 & 261.2 &  769.2 & \nodata & \nodata & \nodata & \nodata & 299.0 &     3.8 &     5.0 & \nodata & \nodata \\
SMP 28  & $-13.57$ & 0.32 & 310.4 &  925.4 &    17.3 &     6.7 &     9.3 &    53.0 & 369.0 &   466.4 &     7.6 &    13.5 & 17.8 \\
SMP 30  & $-13.50$ & 0.11 & 233.4 &  684.7 &    \nodata & \nodata &    \nodata &   291.1 & 311.8 &   843.9 & \nodata &    52.5 & 55.4 \\
SMP 31  & $-12.92$ & 0.54 &  34.9 &  100.0 &     3.4 &     1.0 &     1.2 &    24.5 & 441.2 &    78.1 &     3.6 & \nodata & \nodata \\
SMP 34  & $-12.97$ & 0.06 & 174.1 &  511.1 & \nodata & \nodata & \nodata &     3.4 & 300.0 &    13.1 &     3.8 &     1.0 &  0.9 \\
SMP 46  & $-13.53$ & 0.18 & 396.6 & 1200.0 &    20.9 &     5.1 &     7.4 &    68.5 & 328.1 &   212.2 &     5.7 &    19.5 & 28.6 \\
SMP 53  & $-12.67$ & 0.13 & 401.4 & 1223.3 &     8.8 &     2.9 &     2.8 &    20.3 & 315.8 &    61.9 &     3.7 &     5.5 &  8.7 \\
SMP 58  & $-12.54$ & 0.11 & 228.2 &  666.7 &     2.7 &     1.6 &     0.7 &     2.4 & 312.0 &     8.2 &     4.5 &     0.2 &  0.6 \\
SMP 59
        & $-13.16$ & \nodata & 206.9 &  604.9 & \nodata & \nodata & \nodata &
\nodata &	211\tablenotemark{a} &     \nodata & \nodata &   
	41.8\tablenotemark{b} & \nodata \\
SMP 65  & $-13.44$ & 0.22 & 211.5 &  717.8 & \nodata & \nodata & \nodata & \nodata & 339.7 & \nodata &     4.9 & \nodata & \nodata \\
SMP 71  & $-12.90$ & 0.24 & 392.1 & 1182.5 &    14.5 &     2.7 &     5.3 &    29.9 & 344.4 &    88.9 &     3.8 &     8.1 & 13.2 \\
SMP 78  & $-12.60$ & 0.21 & 456.3 & 1377.0 &     9.6 &     2.6 &     3.4 &    17.4 & 337.3 &    52.0 &     3.6 &     3.1 &  6.5 \\
SMP 79  & $-12.68$ & 0.18 & 421.8 & 1260.7 &     9.3 &     2.6 &     3.4 &     8.6 & 328.4 &    30.5 &     4.1 &     2.6 &  5.1 \\
SMP 80  & $-13.18$ & 0.08 & 175.7 &  533.0 & \nodata & \nodata & \nodata &    10.6 & 304.8 &    29.9 &     4.1 &     3.1 &  4.1 \\
SMP 81  & $-12.66$ & 0.24 & 434.6 & 1304.1 &     4.7 &     3.2 &     1.7 &     5.1 & 347.0 &    18.5 &     4.4 &     0.6 &  1.6 \\
SMP 93
        & $-13.05$ & \nodata & 134.4 &  403.1 &    36.3 & \nodata &    13.2 &
	\nodata &  427\tablenotemark{a}&   \nodata & \nodata &   
	51.4\tablenotemark{b} & \nodata \\
SMP 94\tablenotemark{c}  & $-13.06$ & 1.05 &  18.8 &   52.9 & \nodata & \nodata & \nodata & \nodata & 663.6 & \nodata & \nodata & \nodata & \nodata \\
SMP 95  & $-13.46$ & 0.11 & 335.3 & 1008.7 &    32.4 &    28.8 &    12.9 &    72.6 & 312.0 &   219.0 &    10.9 &    28.9 & 31.2 \\
SMP 100 & $-12.88$ & 0.02 & 391.7 & 1165.4 &     3.1 & \nodata & \nodata &     5.7 & 289.5 &    21.7 & \nodata &     2.4 &  3.5 \\
SMP 102 & $-13.21$ & \tablenotemark{d} & 235.0 &  703.4 & \nodata &     3.8 & \nodata &     1.6 & 228.5 &     4.1 &     2.9 & \nodata & \nodata \\
\enddata  
\tablenotetext{a}{Intensity of the blend of  \ha and \nii 
$\lambda$6548, 6584.}
\tablenotetext{b}{Intensity of the blended doublet 
\sii $\lambda$6716,6731.}
\tablenotetext{c}{Probably not a PN.}
\tablenotetext{d}{Negative extinction.}
\end{deluxetable}


\begin{thebibliography}{}


\bibitem[Balick(1987)]{bal87} Balick, B.\ 1987, \aj, 94, 671


\bibitem[Barlow(1987)]{bar87} Barlow, M.~J.\ 1987, \mnras, 227, 161. 

\bibitem[Boffi \& Stanghellini(1994)]{bs}Boffi, F. R., \& Stanghellini, L. 1994, A\&A, 284, 248


\bibitem[Ciardullo \& Jacoby(1999)]{Ciardullo_Jacoby99}Ciardullo, R., \&
 Jacoby, G.~H. 1999, \apj, 515, 191



\bibitem[Dopita \& Meatheringham(1990)]{dop90}Dopita, M.~A., \&  Meatheringham, S.~J. 1990, ApJ, 357, 140

\bibitem[Meatheringham \& Dopita(1991a)]{dop91a}Meatheringham, S.~J., \& Dopita, M.~A. 1991a, ApJS, 75, 407

\bibitem[Meatheringham \& Dopita(1991b)]{dop91b}Meatheringham, S.~J., \& Dopita, M.~A. 1991b, ApJS, 76, 1085



\bibitem[Frank(2000)]{fra00} Frank, A.\ 2000, ASP 
Conf.~Ser.~199: Asymmetrical Planetary Nebulae II: From Origins to 
Microstructures, 225

\bibitem[Icke, Preston \& Balick(1989)]{ipb89}Icke, V., Preston, H.~L., \& Balick, B.\ 1989, \aj, 97, 462


\bibitem[Howarth (1983)]{how}Howarth, I. D. 1983, MNRAS, 203, 301


\bibitem[Leitherer \& Bohlin(1997)]{Leitherer_Bohlin97}Leitherer, C., \& 
	Bohlin, R.~C. 1997, STIS Instrument Science Report 97--13 
	(Baltimore: ST~ScI)
	
\bibitem[Leitherer, et al.(2000)]{Leitherer_etal00}Leitherer, C., 
	et al.~2000, STIS Instrument Handbook, Version 4.0 (Baltimore: ST~ScI)


\bibitem[McGrath, Busko, \& Hodge(1999)]{McGrath_etal99}McGrath, M.~A., 
	Busko, I., \& Hodge, P.~E.~1999, STIS Instrument Science Report 
	99--03 (Baltimore: ST~ScI)

\bibitem[Manchado et al.(1996a)]{man96a}Manchado, A., Guerrero, M., Stanghellini, L., \&
Serra-Ricart, M. 1996a, 
The IAC Morphological Catalog of Northern Galactic Planetary
Nebulae (Tenerife: Instituto de Astrof\'{\i}sica de Canarias)

\bibitem[Manchado, Stanghellini, \& Guerrero(1996b)]{man96b}
Manchado, A., Stanghellini, L., \& Guerrero, M.~A.\ 1996b, \apjl, 466, L95

\bibitem[Manchado et al.(2000)]{man00}Manchado, A., Villaver, E., 
Stanghellini, L., \& Guerrero, M. A.\ 2000, ASP Conf.\ Ser.\ 199: 
Asymmetrical Planetary Nebulae II: From Origins to Microstructures, 17 


\bibitem[Osterbrock(1989)]{ost89}Osterbrock, D. E. 1989, Astrophysics of
Gaseous Nebulae and Active Galactic Nuclei 
(Mill Valley, CA: University Science Books)

\bibitem[Peimbert(1978)]{pei78}Peimbert, M. 1978, IAU Symp. 76, ed. Y. Terzian,
Reidel, 215

\bibitem[Torres-Peimbert \& Peimbert(1997)]{pei97}Torres-Peimbert, S., \& Peimbert, M. 1997, IAU Symp. 180, eds.
H. Habing and H. Lamers, Kluwer, 175


\bibitem[Savage \& Mathis (1979)]{Savage79} Savage, B., \& Mathis, J. 1979, 
	\araa, 17, 73

\bibitem[Shaw \& Dufour (1995)]{sha95}Shaw, R.~A.~\& Dufour, R.~J.\ 1995, \pasp, 107, 896	
	
\bibitem[Shaw et al.(2001)]{sha01} Shaw, R.~A., Stanghellini, L., 
Mutchler, M., Balick, B., \& Blades, J.~C.\ 2001, \apj, 548, 727 (Paper I)


\bibitem[Stanghellini \& Kaler(1989)]{sk89}Stanghellini, L., \& Kaler, J. B. 1989, ApJ, 343, 811 

\bibitem[Stanghellini, Corradi, \& Schwarz(1993)]{sta93} Stanghellini, L., Corradi, 
R.~L.~M., \& Schwarz, H.~E. 1993, A\&A, 279, 521

\bibitem[Stanghellini et al.(1999)]{sta99}Stanghellini, L., Blades, J.~C., 
Osmer, S.~J., Barlow, M.~J., \& Liu, X.-W. 1999, ApJ, 510, 687 


\bibitem[Stanghellini, et al.(2000)]{sta00}Stanghellini, L., 
	Shaw, R.~A., Balick, B., \& Blades, J.~C.~2000, \apjl, 534, L167
	

\bibitem[Stys \& Walborn(2001)]{Stys_Walborn01}Stys, D.~J., \& Walborn, N.~R. 
	2001, STIS Instrument Science Report 2001--01R (Baltimore: ST~ScI)


\bibitem[van den Hoek \& Groenewegen(1997)]{van97}van den Hoek, L. B., 
\& Groenewegen, M.~A.~T. 1997, A\&AS, 123, 305

\bibitem[Vassiliadis et al.(1992)] {vdm} Vassiliadis, E., Dopita, M. A., Morgan, 
D. H., \& Bell, J. F. 1992, ApJS, 83, 87



\end{thebibliography}
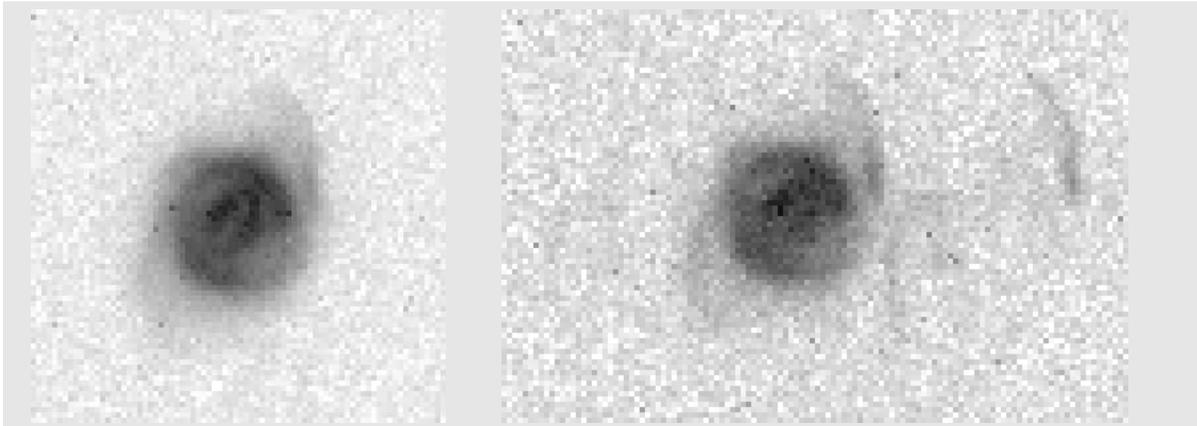
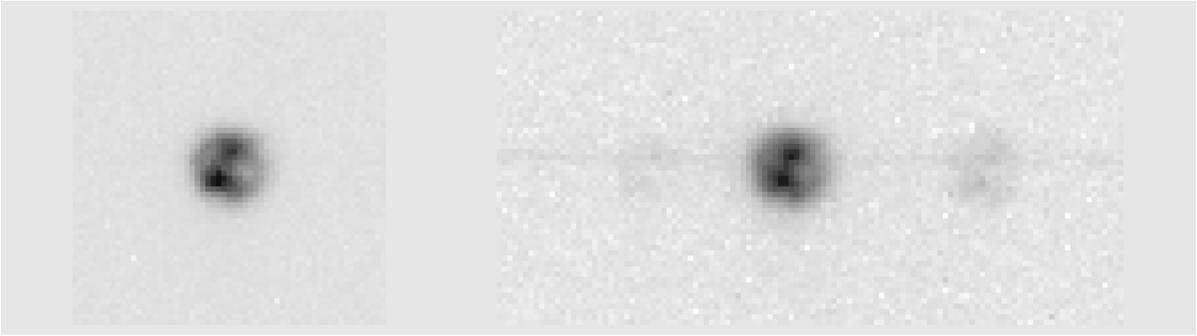
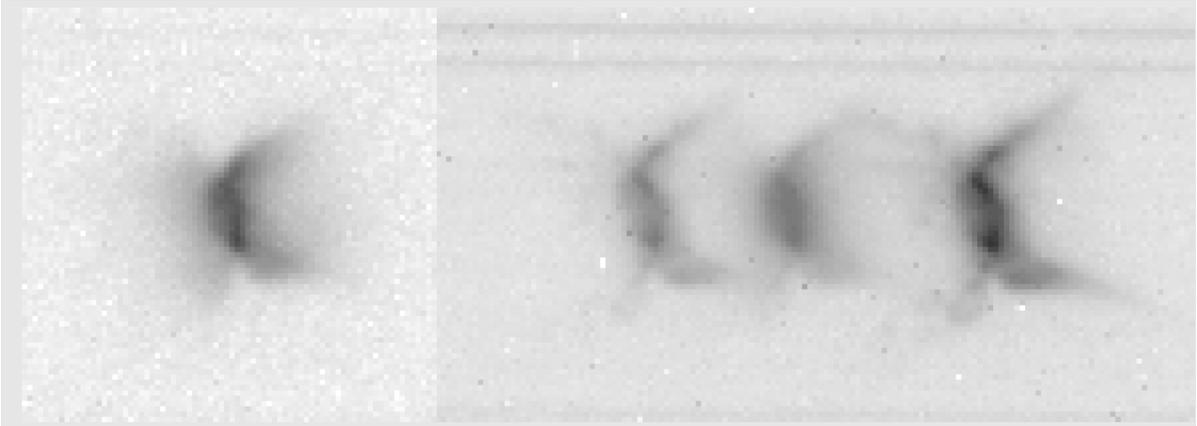
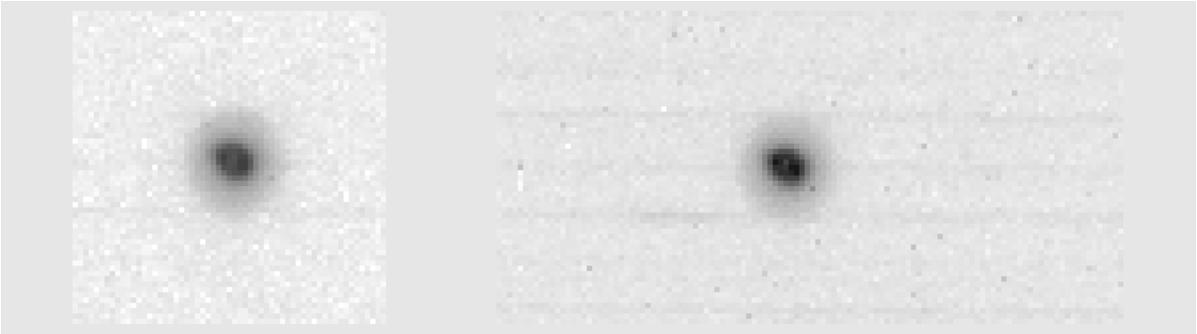
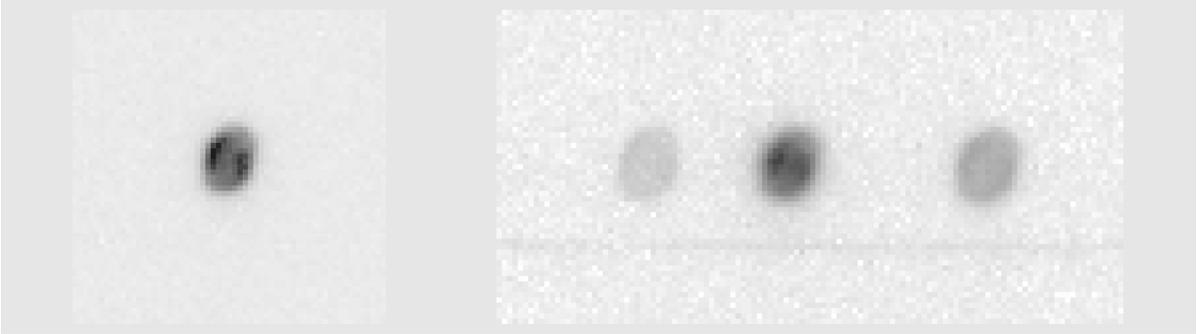
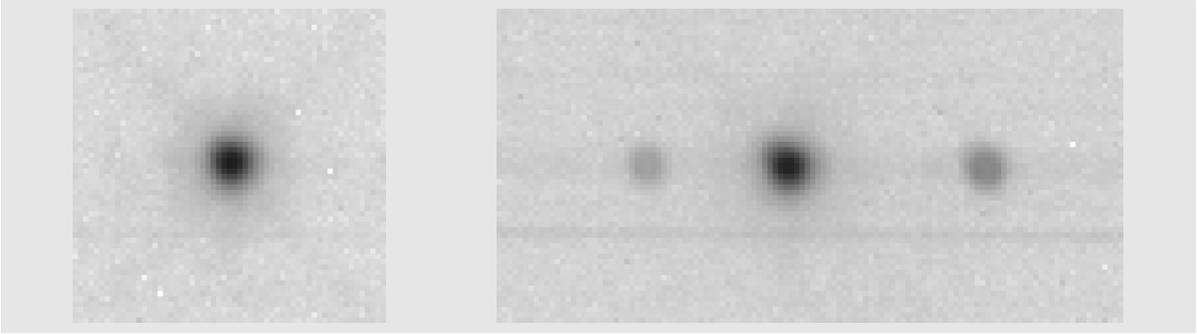
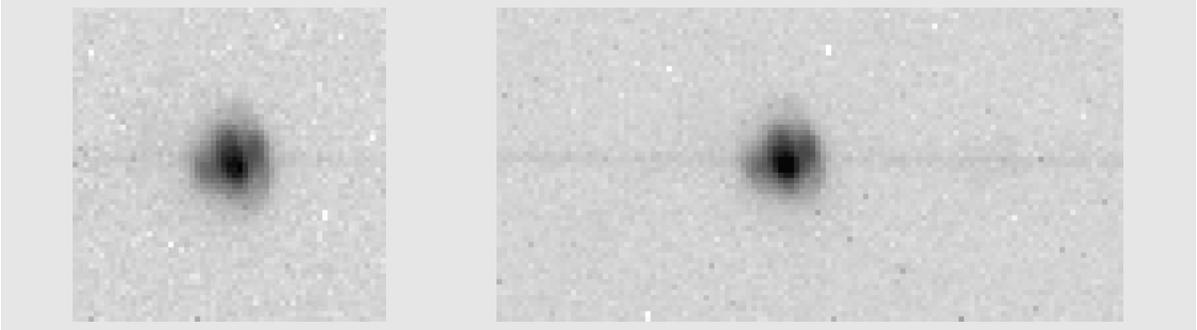
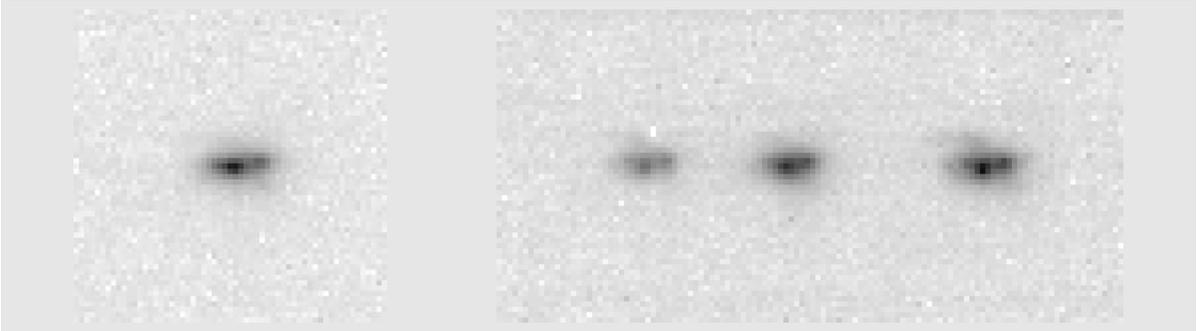
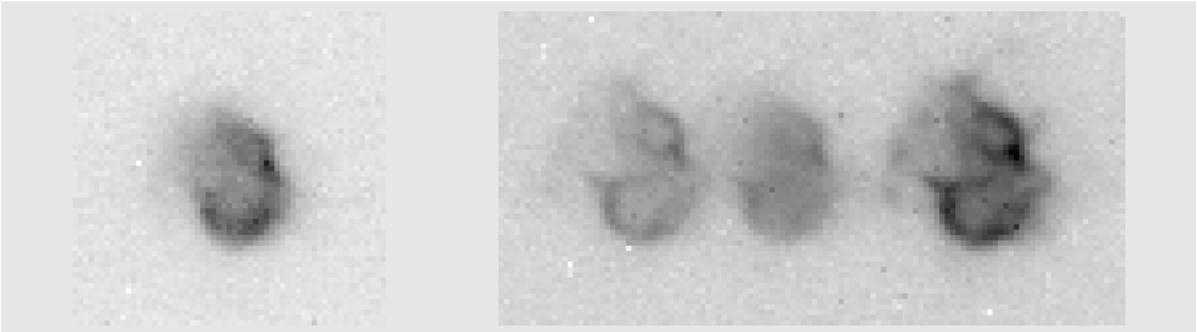
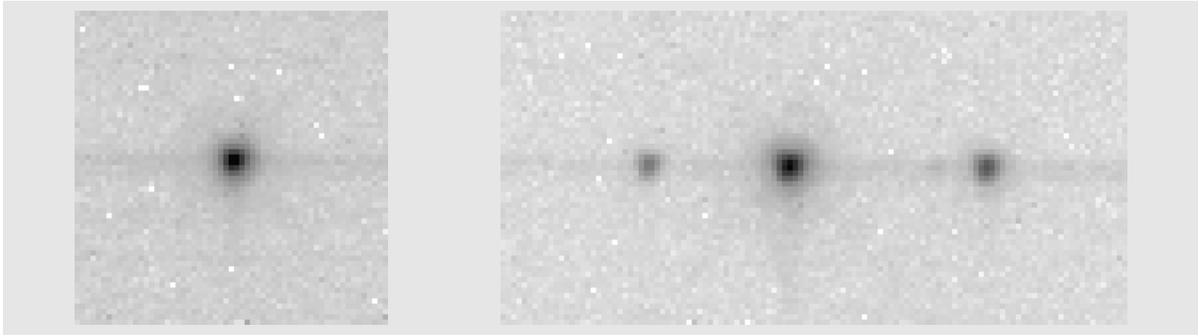
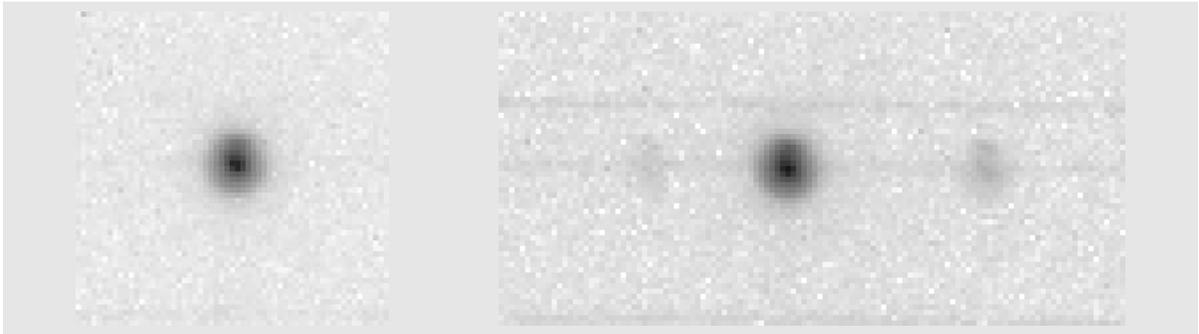
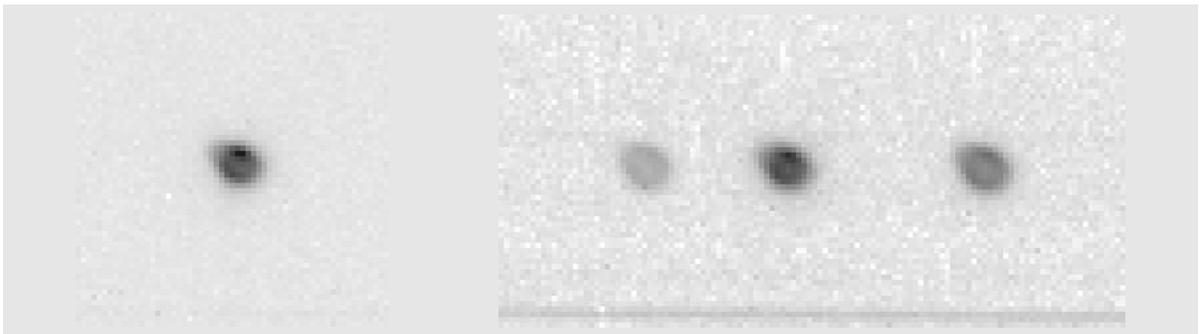
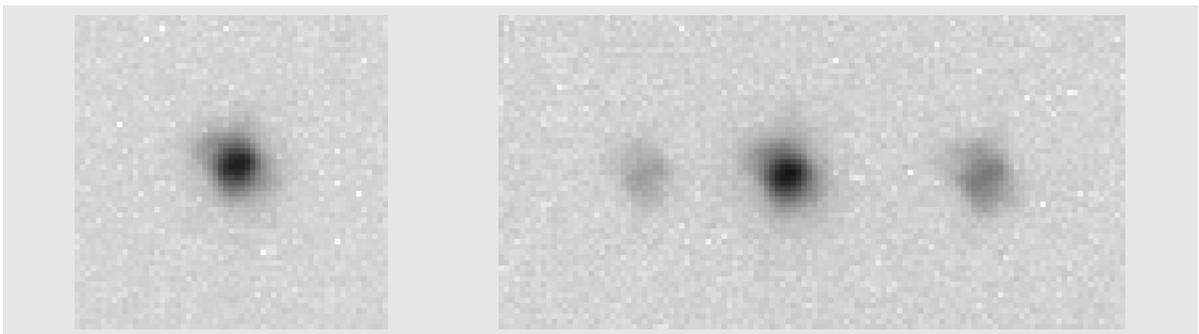
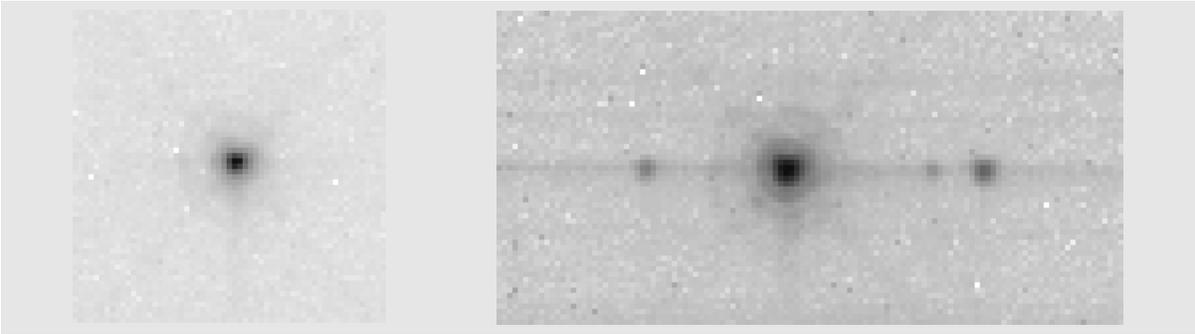
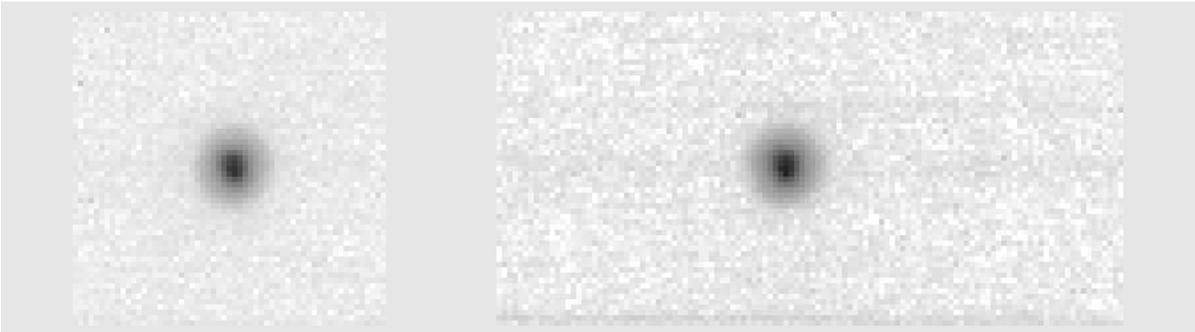
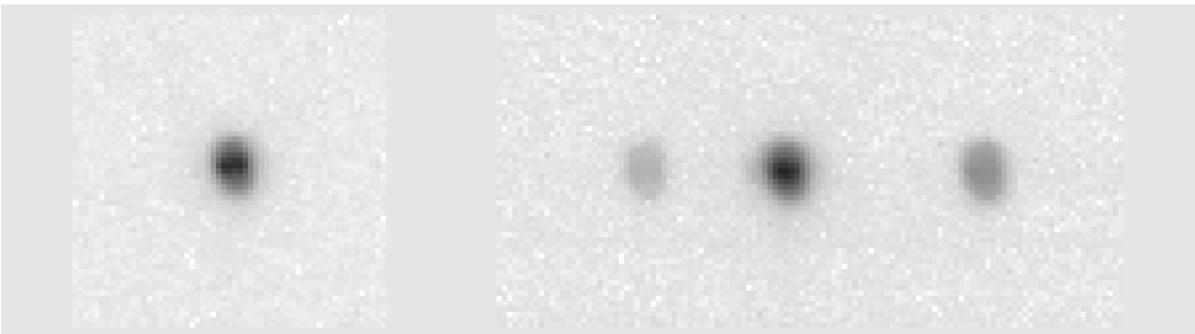
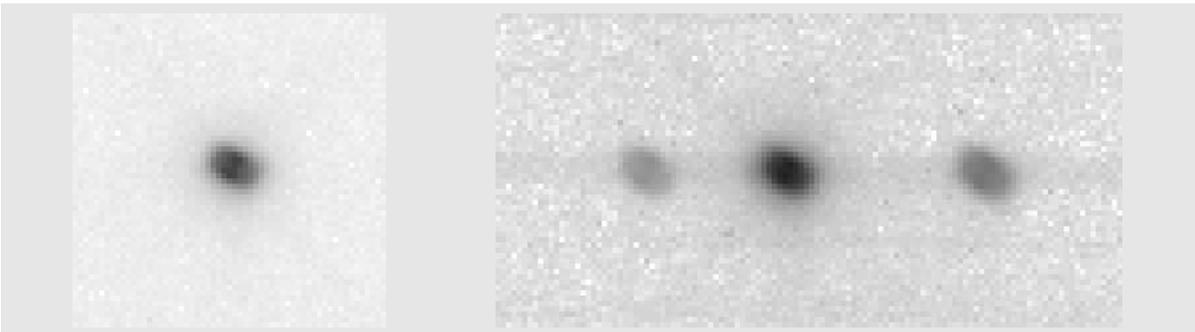
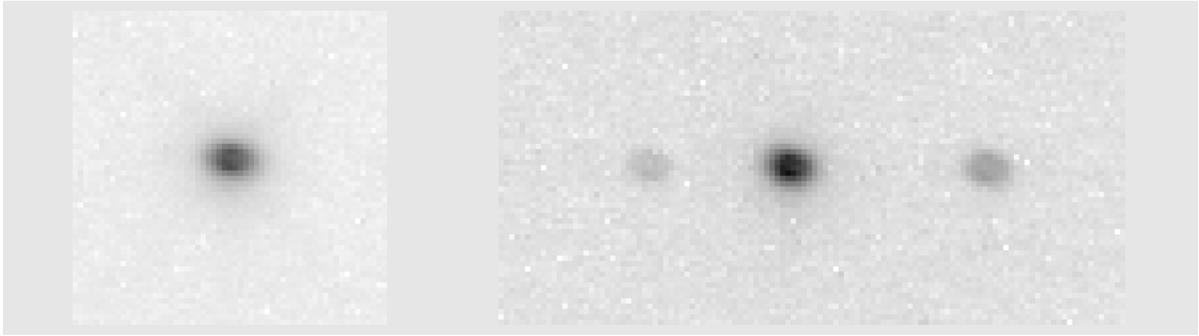
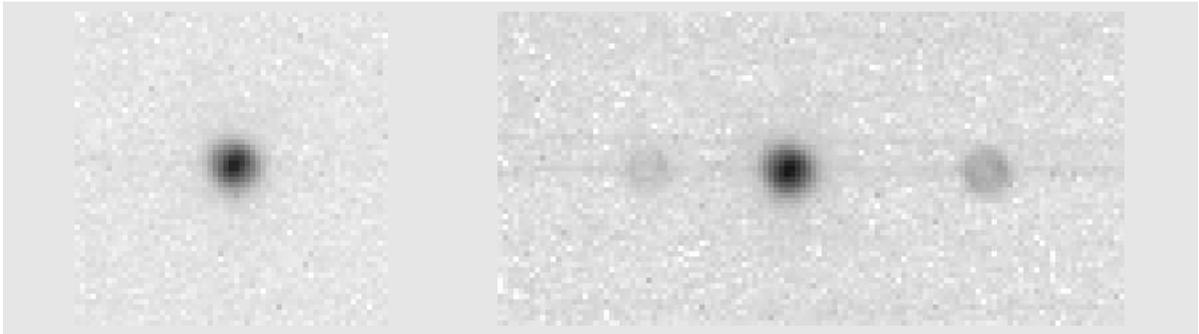
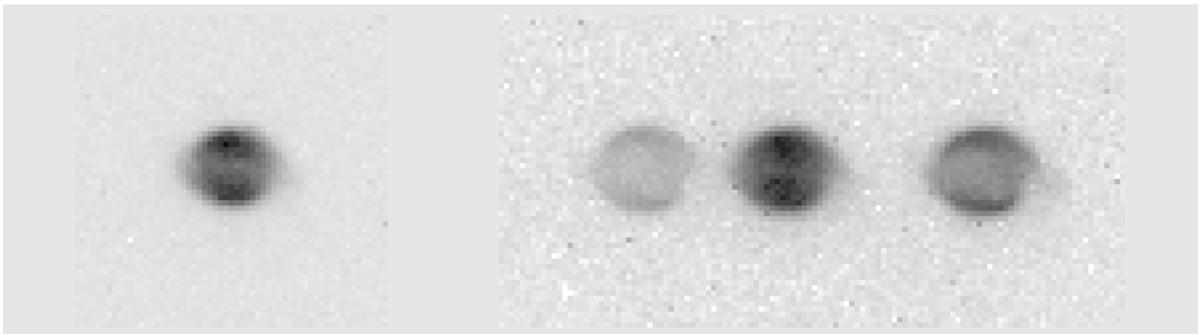
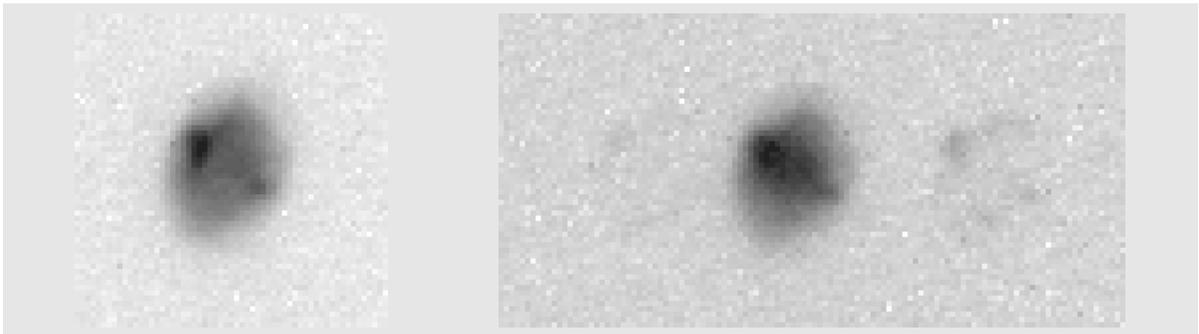
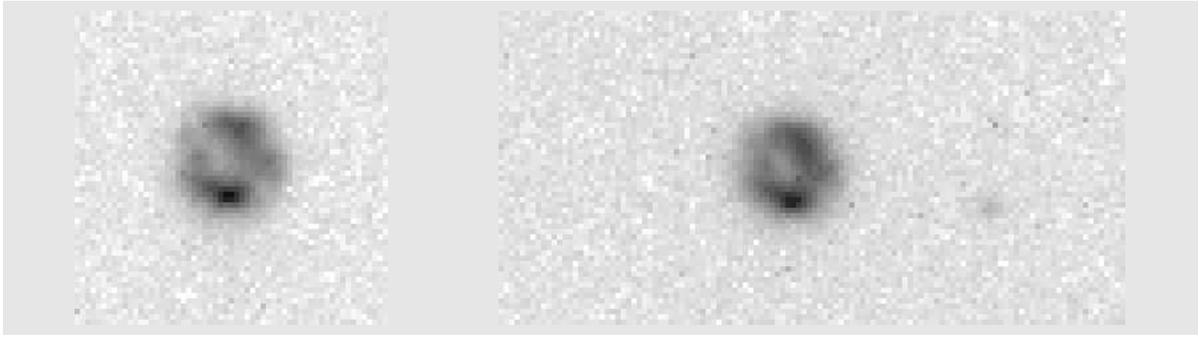
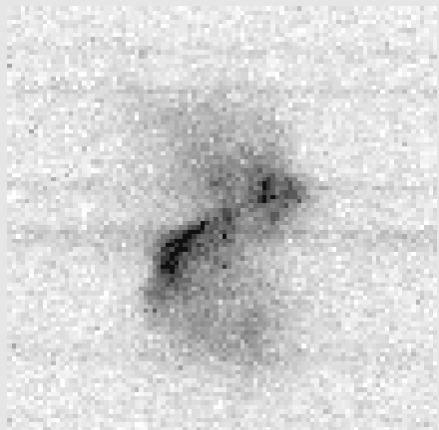
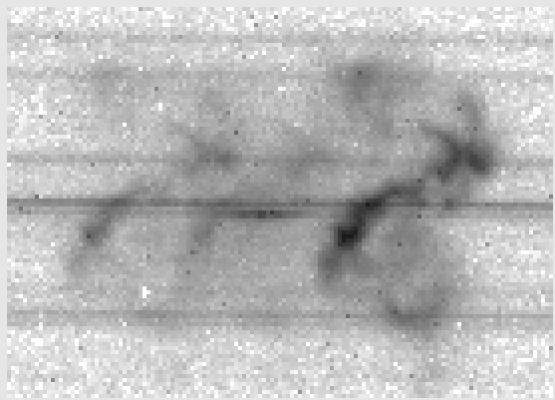
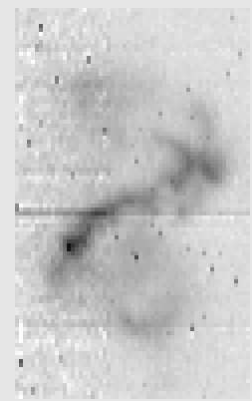
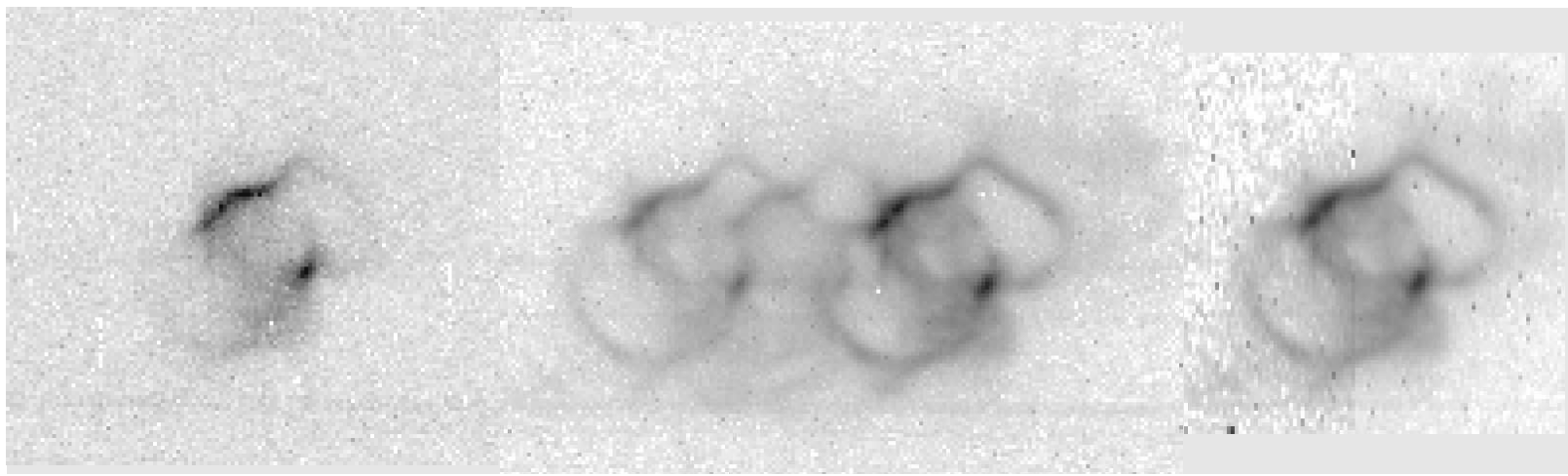
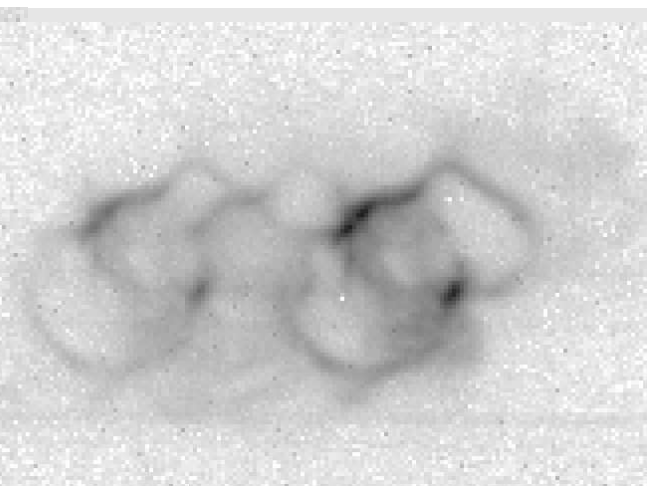
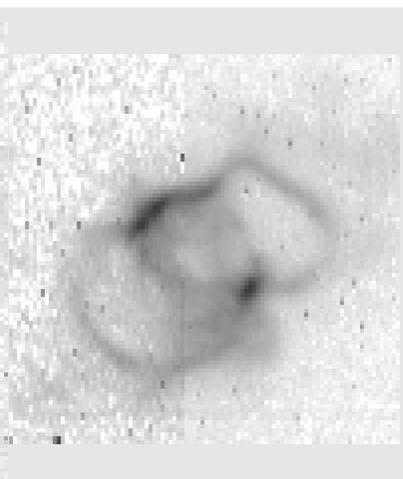
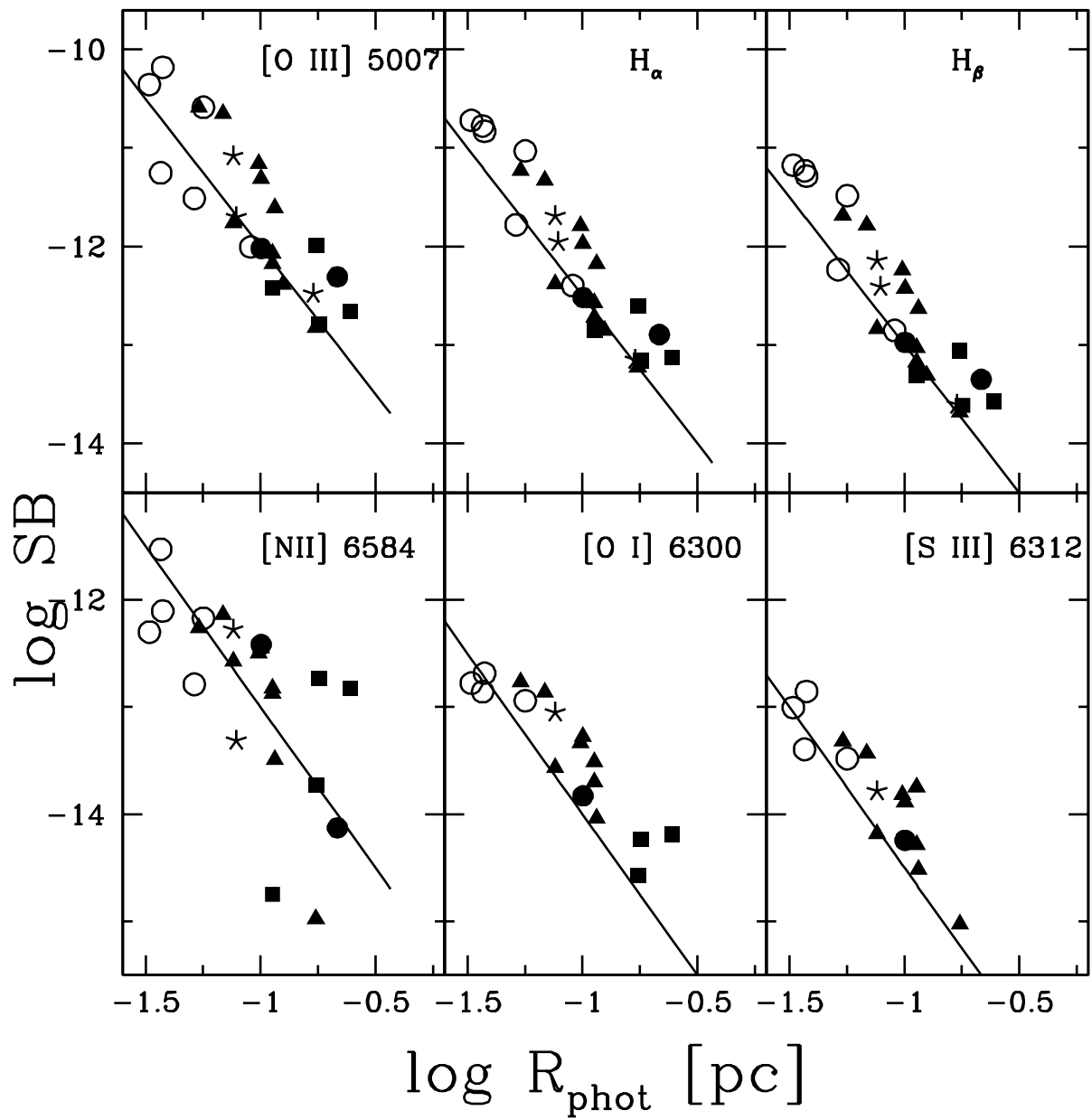
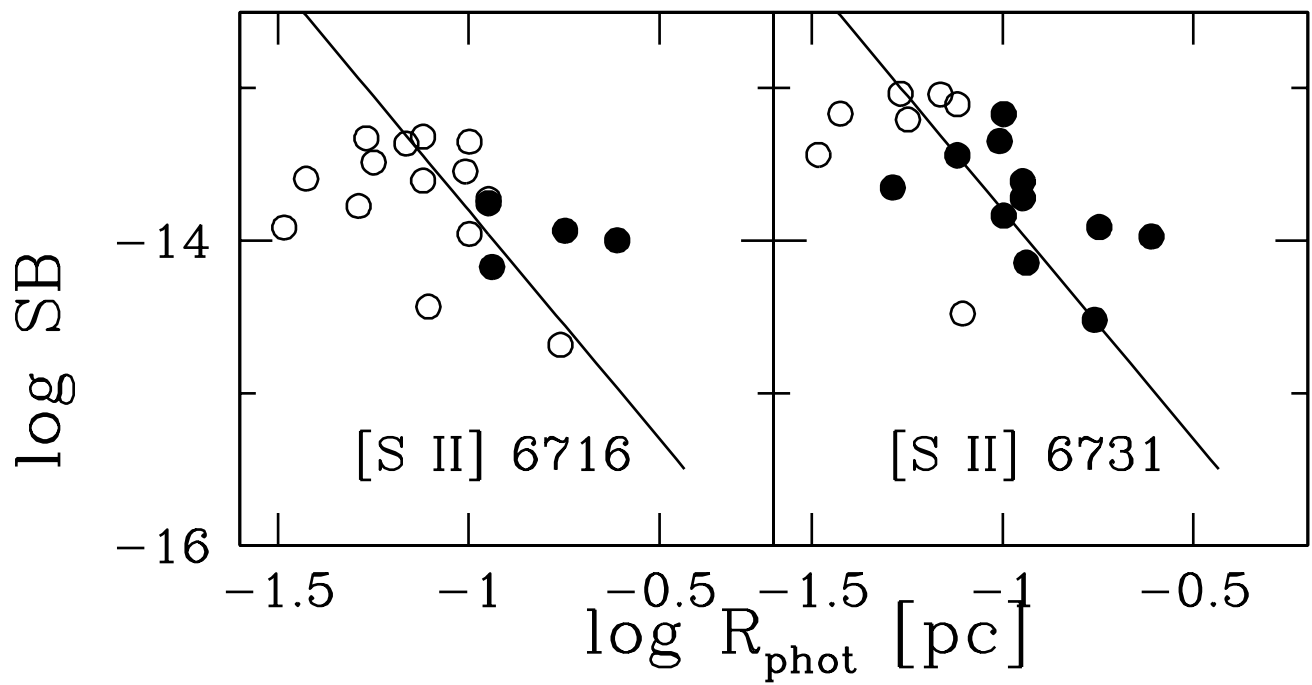
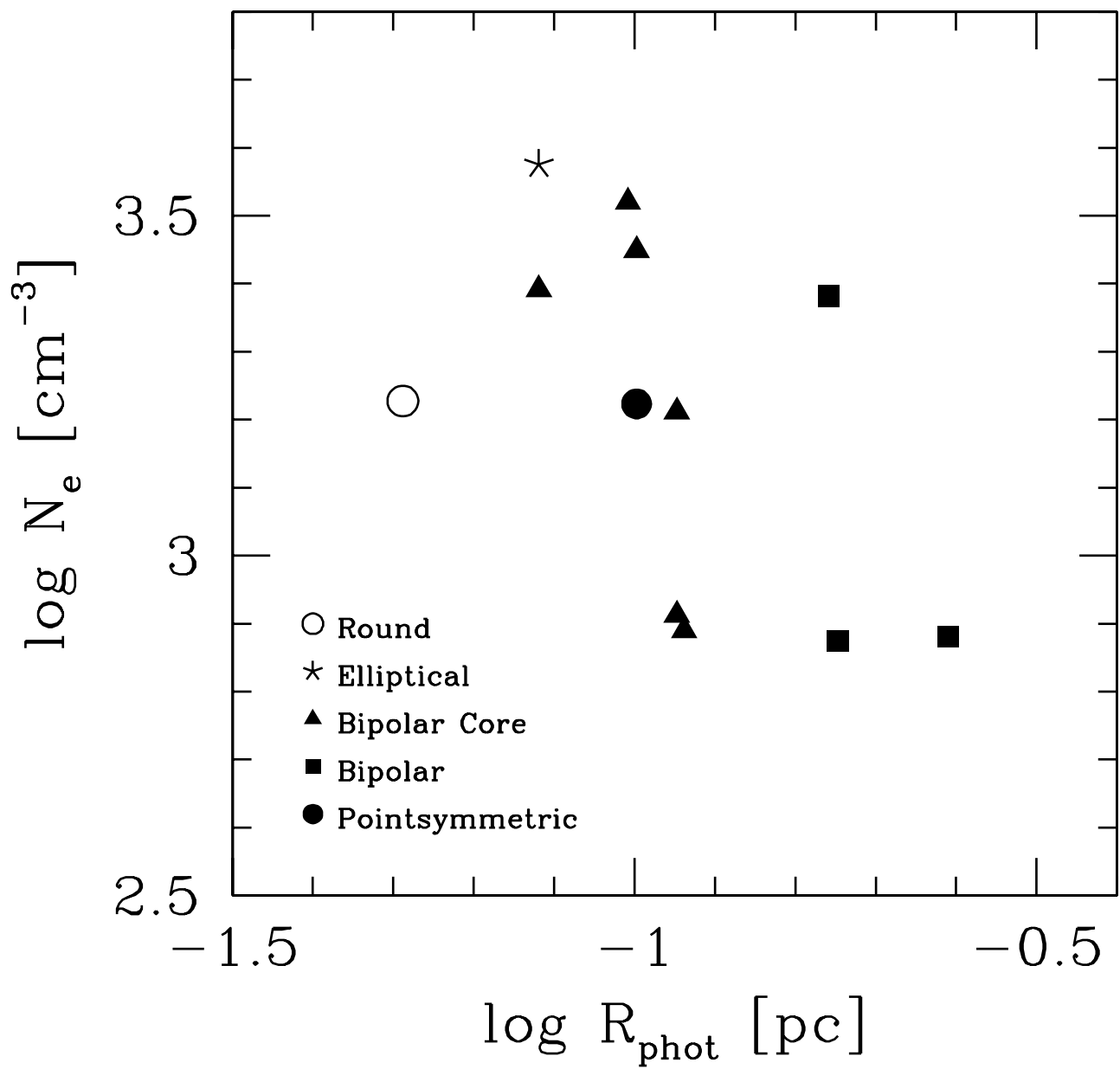
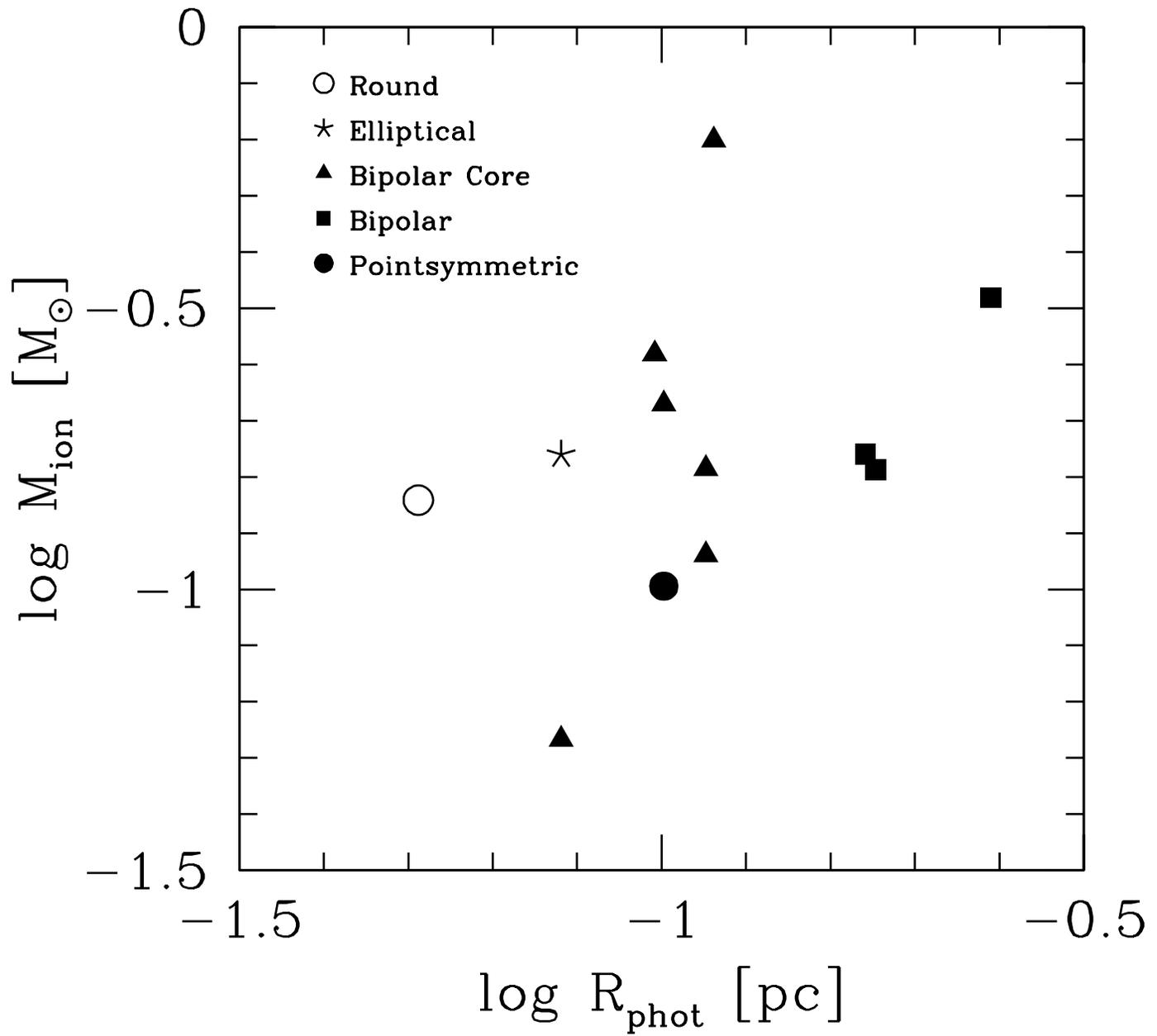
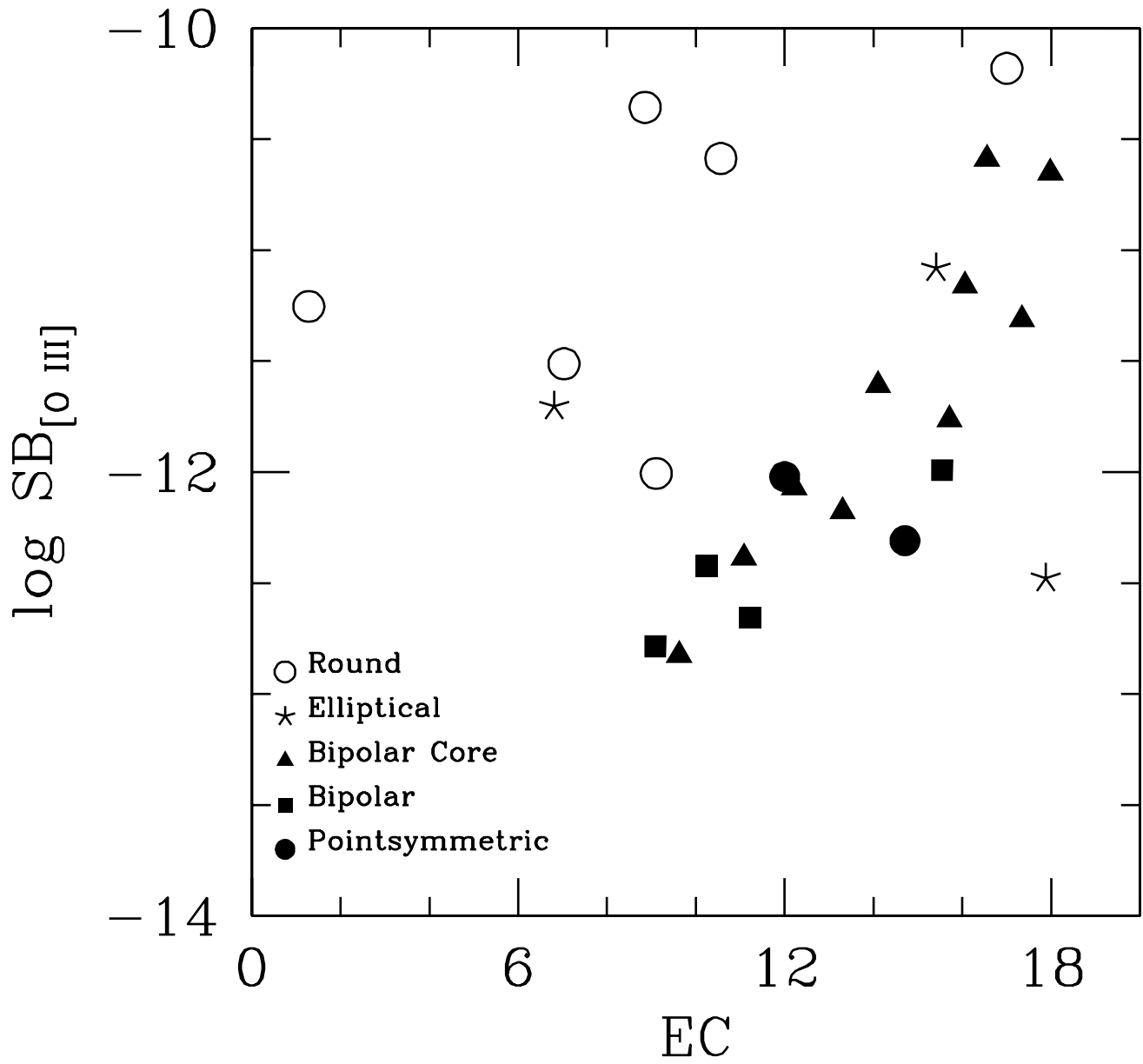
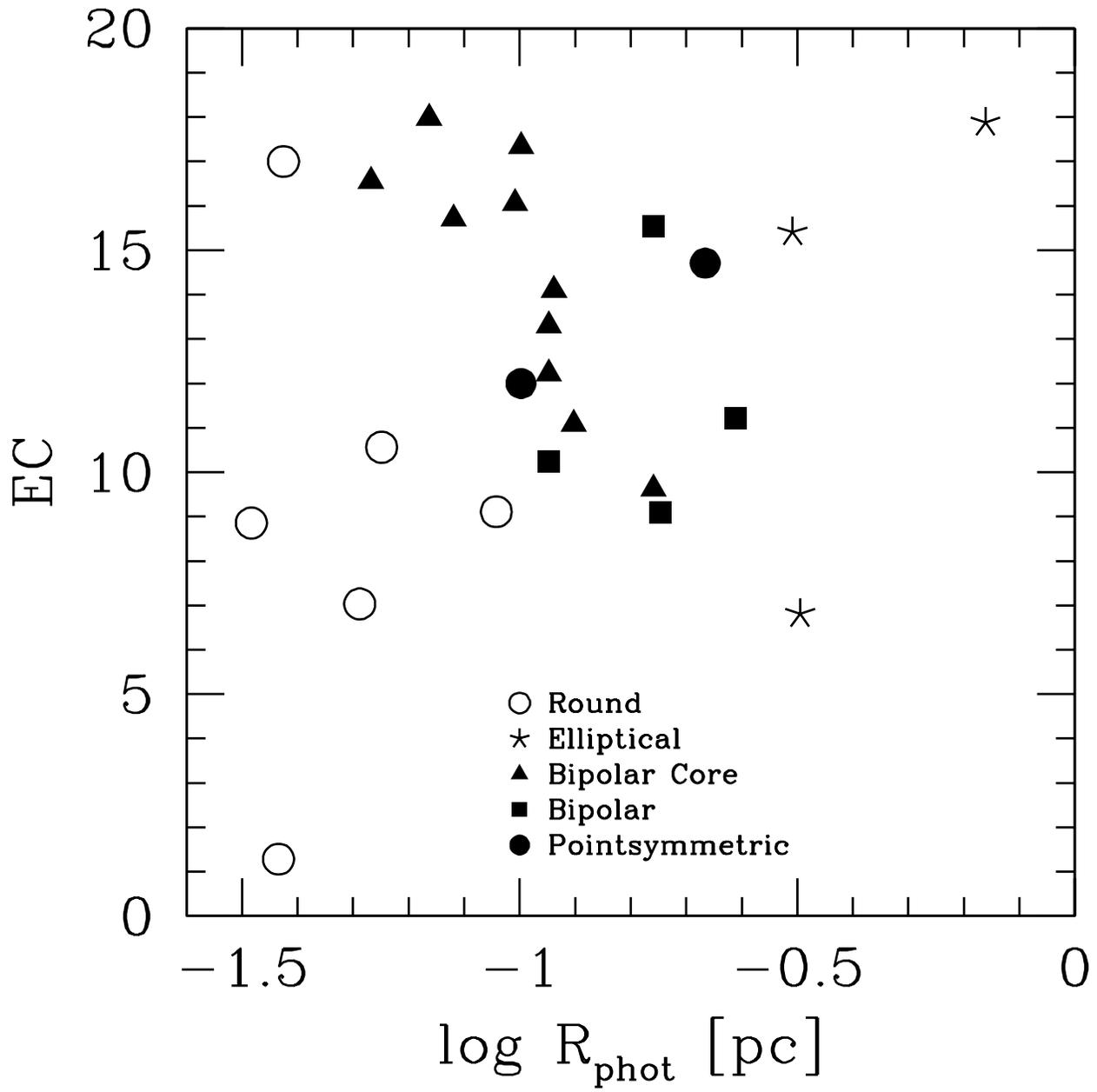
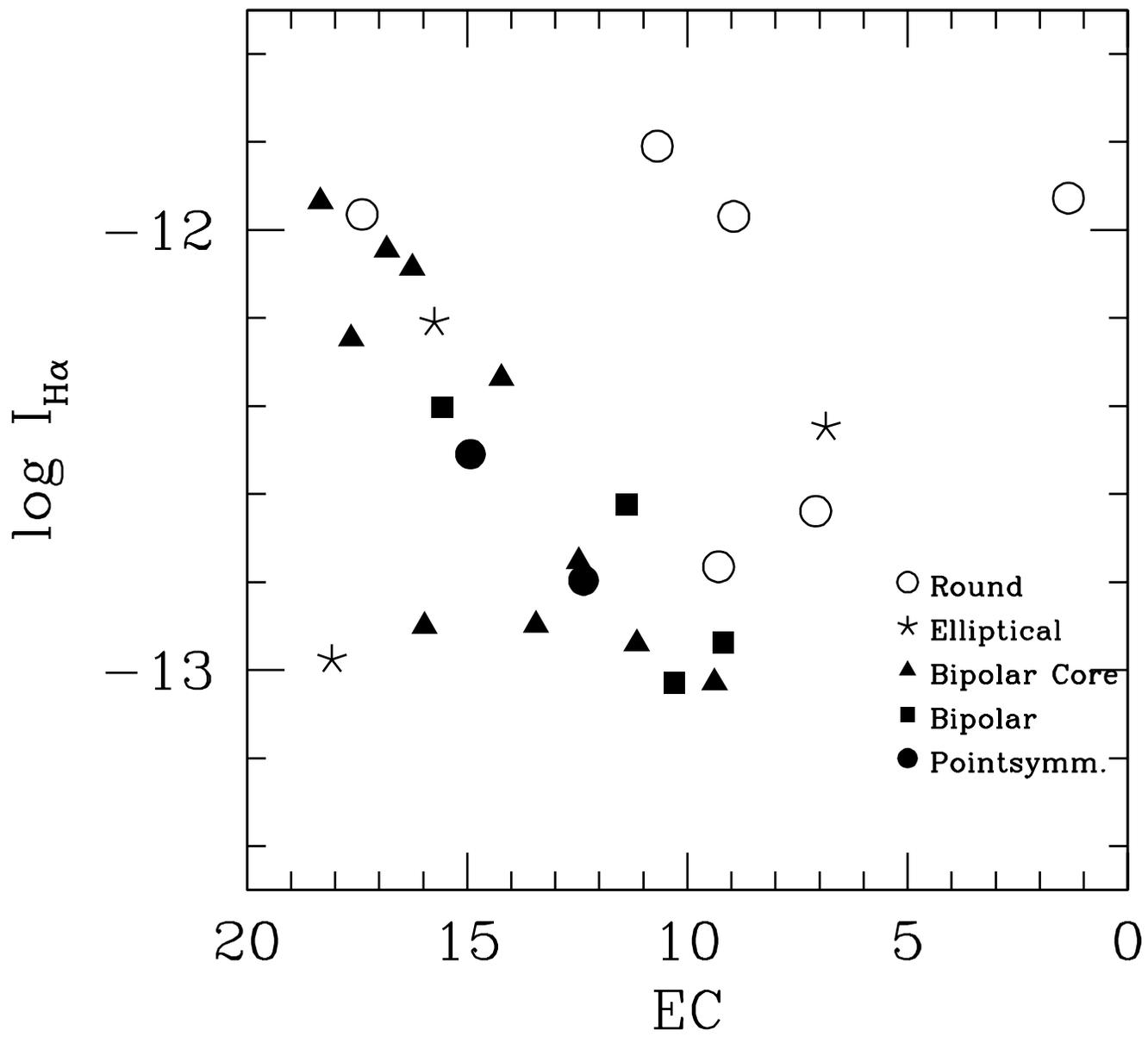
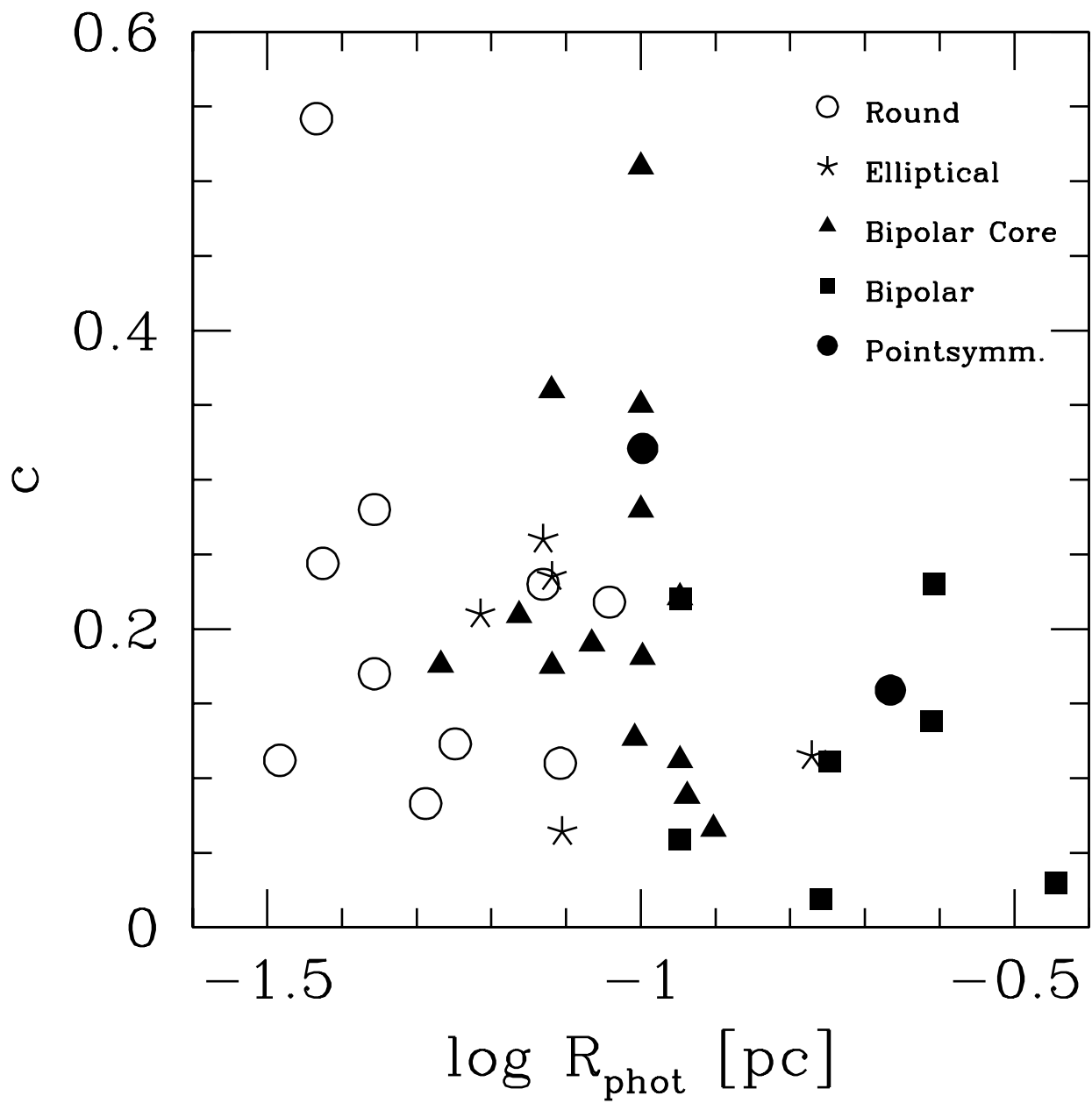
\end{document}